\documentclass[twocolumn,prb,preprintnumbers,amsmath,amssymb,letterpaper,floatfix,showpacs]{revtex4}
\usepackage[utf8]{inputenc}
\usepackage{gensymb}
\usepackage{upgreek}
\usepackage{float} 
\usepackage{graphicx} 

\begin{document}

\title{Chemical Substitution and High Pressure Effects on Superconductors in the \textit{Ln}OBiS$_{2}$ (\textit{Ln} = La-Nd) System}

\author{Y. Fang$^{1,3}$}
\author{C. T. Wolowiec$^{2,3}$}
\author{D. Yazici$^{2,3}$}
\author{M. B. Maple$^{1,2,3}$}

\affiliation{$^1$Materials Science and Engineering Program, University of California, San Diego, La Jolla, California 92093, USA}
\affiliation{$^2$Department of Physics, University of California, San Diego, La Jolla, California 92093, USA}
\affiliation{$^3$Center for Advanced Nanoscience, University of California, San Diego, La Jolla, California 92093, USA}

\email[Corresponding Author: ]{mbmaple@ucsd.edu}

  \begin{abstract}
{A large number of compounds which contain BiS$_{2}$ layers exhibit enhanced superconductivity upon electron doping. Much interest and research effort has been focused on BiS$_{2}$-based compounds which provide new opportunities for exploring the nature of superconductivity. Important to the study of BiS$_{2}$-based superconductors is the relation between structure and superconductivity.  By modifying either the superconducting BiS$_2$ layers or the blocking layers in these layered compounds, one can effectively tune the lattice parameters, local atomic environment, electronic structure, and other physical properties of these materials.  In this article, we will review some of the recent progress on research of the effects of chemical substitution in  BiS$_{2}$-based compounds, with special attention given to the compounds in the \textit{Ln}OBiS$_{2}$ (\textit{Ln} = La-Nd) system. Strategies which are reported to be essential in optimizing superconductivity of these materials will also be discussed.}
\end{abstract}
\pacs{74.25.F-, 74.25.Dw, 74.62.Bf}

\maketitle
\section{Introduction}

Following the discovery of superconductivity by Kammerlingh Onnes in 1911, efforts in the search for new superconducting materials during the past century have shown that most of the pure metals and an enormous number of alloys and compounds become superconducting at low temperatures. The highest superconducting critical temperature \textit{T$_c$} that has been reported is 133 K for the cuprates at ambient pressure and 203 K in sulfur hydride at 155 GPa, although the latter claim has not yet been independently verified \cite{bednorz1986possible,drozdov2015conventional}. In order to experimentally realize new superconducting materials with properties suitable for practical applications, there is still much to accomplish in both theory and  experiment. The recently discovered family of BiS$_{2}$-based superconductors exhibit a layered crystal structure which is characterized by an alternating pattern of BiS$_{2}$ double layers that are separated by blocking layers \cite{mizuguchi2012bis2}. This layered structure provides new hope and opportunity to enrich the class of layered superconductors with higher values of $T_c$.

Chemical substitution may induce significant change in materials to their lattice parameters, crystal structure and electronic structure. It has been broadly used as a strategy to tune physical properties of materials including electronic transport, magnetic, thermal, mechanical, and luminescent properties. Chemical substitution has been found to induce superconductivity via electron or hole doping in $Ln$FeAsO compounds by replacing F for O \cite{kamihara_2008_1, chen_2008_1, ren_2008_1, ren_2008_3, ren_2008_2, fang_2008_2, xing2012superconductivity}, Co for Fe \cite{sales_2008_2}, Sr for La \cite{zhu_2008_1}, Th for Gd \cite{ren_2008_5}, and also through the introduction of oxygen vacancies \cite{ren_2008_6,yang_2008_1}. Similar to the $Ln$FeAsO compounds, it has been reported that electron carriers are essential for the emergence of superconductivity in the BiS$_2$-based materials \cite{mizuguchi2012bis2,yildirim2013ferroelectric}. By modifying the blocking layers of the BiS$_{2}$-based compounds, electron carriers can be generated within the structures. As a result, a large number of novel BiS$_{2}$-based superconductors, such as Bi$_4$O$_4$S$_3$, Bi$_3$O$_2$S$_3$, \textit{Ln}O$_{1-x}$F$_{x}$BiS$_{2}$ (\textit{Ln} = La, Ce, Pr, Nd, Yb), La$_{1-x}$$M$$_{x}$OBiS$_{2}$ ($M$ = Ti, Zr, Hf, Th), Sr$_{1-x}$La$_{x}$FBiS$_{2}$, EuBiS$_2$F, and Eu$_{3}$Bi$_{2}$S$_{4}$F$_{4}$, have been developed with $T_c$'s ranging from 2.7 to 10.6 K \cite{mizuguchi2012bis2,shao2014bulk,awana2013appearance,demura2013new,mizuguchi2012superconductivity,xing2012superconductivity,yazici2013superconductivity,mizuguchi2014superconductivity,zhai2014possible,deguchi2013evolution,yazici2013superconductivity2,lin2013superconductivity}.

The superconducting properties of BiS$_{2}$-based compounds are also sensitive to the environment in which they are synthesized or being measured. For example, polycrystalline samples of LaO$_{1-x}$F$_{x}$BiS$_{2}$ subjected to high pressure (HP) annealing have $T_c$ values significantly higher than as-grown (AG) samples. Also, with the application of external hydrostatic pressure, a monotonic decrease in $T_c$ is observed for both the Bi$_{4}$O$_{4}$S$_{3}$ and SrFBiS$_{2}$ compounds \cite{kotegawa2012pressure,jha2014hydrostatic}; in contrast, the  compounds \textit{Ln}O$_{1-x}$F$_{x}$BiS$_{2}$, Eu$_{3}$Bi$_{2}$S$_{4}$F$_{4}$, EuFBiS$_{2}$, LaO$_{0.5}$F$_{0.5}$BiSe$_{2}$ and Sr$_{0.5}$La$_{0.5}$FBiS$_{2}$ were reported to have a sharp jump in $T_c$ under applied pressure, indicating a pressure-induced transition from a low-$T_c$ to high-$T_c$ superconductor phase \cite{wolowiec2013,wolowiec2013pressure,kotegawa2012pressure,Jha,guo2015evidence,luo2014pressure,fujioka2014pressure}. It is also observed that superconductivity in the BiS$_{2}$-based compounds is very sensitive to chemical composition and to the level of electron doping in particular. As a result, $T_c$ values of the samples with the same nominal chemical composition are not always the same since there might be slight differences in the actual chemical composition.

At present, there have been more than one hundred research papers published regarding superconductivity in the BiS$_{2}$-based compounds, which have helped increase our understanding of these novel materials and superconductivity in general. The peculiar crystal structure of these BiS$_2$-based compounds as a landscape for studying and optimizing superconductivity has fostered a healthy competition within the scientific community to explore the important critical parameters that affect $T_c$ as well as to understand the nature of superconductivity itself in these layered structures. In this article, we will review some of the recent progress in both the study of chemical substitution and external pressure effects on the BiS$_{2}$-based compounds. Discussions about some of the parameters that are essential for superconductivity in these compounds will also be presented.

%

\section{Bi$_{4}$O$_{4}$S$_{3}$}

As the first discovered superconductor in the BiS$_{2}$-based compounds, Bi$_{4}$O$_{4}$S$_{3}$ is a bulk superconductor with a $T_c$ of about 4.5 K \cite{mizuguchi2012bis2,singh2012bulk}. The structure of Bi$_{4}$O$_{4}$S$_{3}$ is derived from the structure of the parent Bi$_{6}$O$_{8}$S$_{5}$ with a 50\% deficiency in SO$_{4}$ units; this results in the presence of electron carriers in the BiS$_2$ layers which induce the superconductivity in Bi$_{4}$O$_{4}$S$_{3}$ \cite{mizuguchi2014review,mizuguchi2012bis2}. Interestingly, the parent phase, Bi$_{6}$O$_{8}$S$_{5}$, which was initially regarded as a band insulator, was soon found to show superconductivity at $\sim$4.2 K when the sample is annealed in vacuum \cite{yu2013superconductivity}.

Chemical substitutions have been found to suppress $T_c$ in the superconducting compound Bi$_{4}$O$_{4}$S$_{3}$. Substitution of Se at the S site results in a slight increase in the lattice parameter $c$ and a remarkable enhancement of the normal state resistivity \cite{jha2014effect}. The $T_c$ of the Bi$_4$O$_4$S$_{3-x}$Se$_x$ compound decreases with increasing Se concentration from 4.4 K at $x$ = 0 to 3.5 K at $x$ = 0.15, while the upper critical field at 0 K, $H_{c2}$(0), determined by fitting the conventional one-band Werthamer–Helfand–Hohenberg (WHH) equation to the $H_{c2}$($T$) data, is also suppressed from 2.3 to 1.8 T \cite{WHH}. Similarly, $T_c$ of the Bi$_{4-x}$Ag$_{x}$O$_{4}$S$_{3}$ compound gradually decreases with increasing Ag concentration and a subtle shrinkage of both the $a$ and $c$ lattice parameters were observed up to $x$ = 0.2 \cite{tan2012suppression}. It should be noted that the chemical pressure generated by Ag substitution is negligible in Bi$_{4}$O$_{4}$S$_{3}$ and the suppression of $T_c$ most likely arises from the substitution of the Ag ions into the superconducting BiS$_{2}$ layers and the resultant change in the electronic structure of the compound.

\section{\textit{Ln}O$_{1-x}$F$_{x}$BiS$_{2}$ (\textit{Ln} = La, Ce, Pr, Nd, Yb)}

The parent compounds in the \textit{Ln}OBiS$_{2}$ family are band insulators. The substitution of F for O induces superconductivity via electron doping into the BiS$_{2}$ layers. By applying relatively small magnetic fields ($\sim$1 T) to polycrysatlline specimens of these superconductors, the superconducting state can be destroyed \cite{yazici2013superconductivity}. Due to the chemical similarity of the rare earth elements, there is considerable flexibility in modifications that can be made to the blocking layers of these compounds which facilitates the investigation of the superconducting and normal state properties of BiS$_2$-based compounds. The \textit{Ln}OBiS$_{2}$ compounds crystallize in a tetragonal structure with space group \textit{P4/nmm} \cite{mizuguchi2012superconductivity}. Between each \textit{Ln}O blocking layer, there is a double BiS$_{2}$ superconducting layer in which there are two distinct positions for the S ions: in-plane position (S1) and out-of-plane position (S2), as indicated in Fig. 1.

\begin{figure}[t]
\centering
\includegraphics[width=8cm]{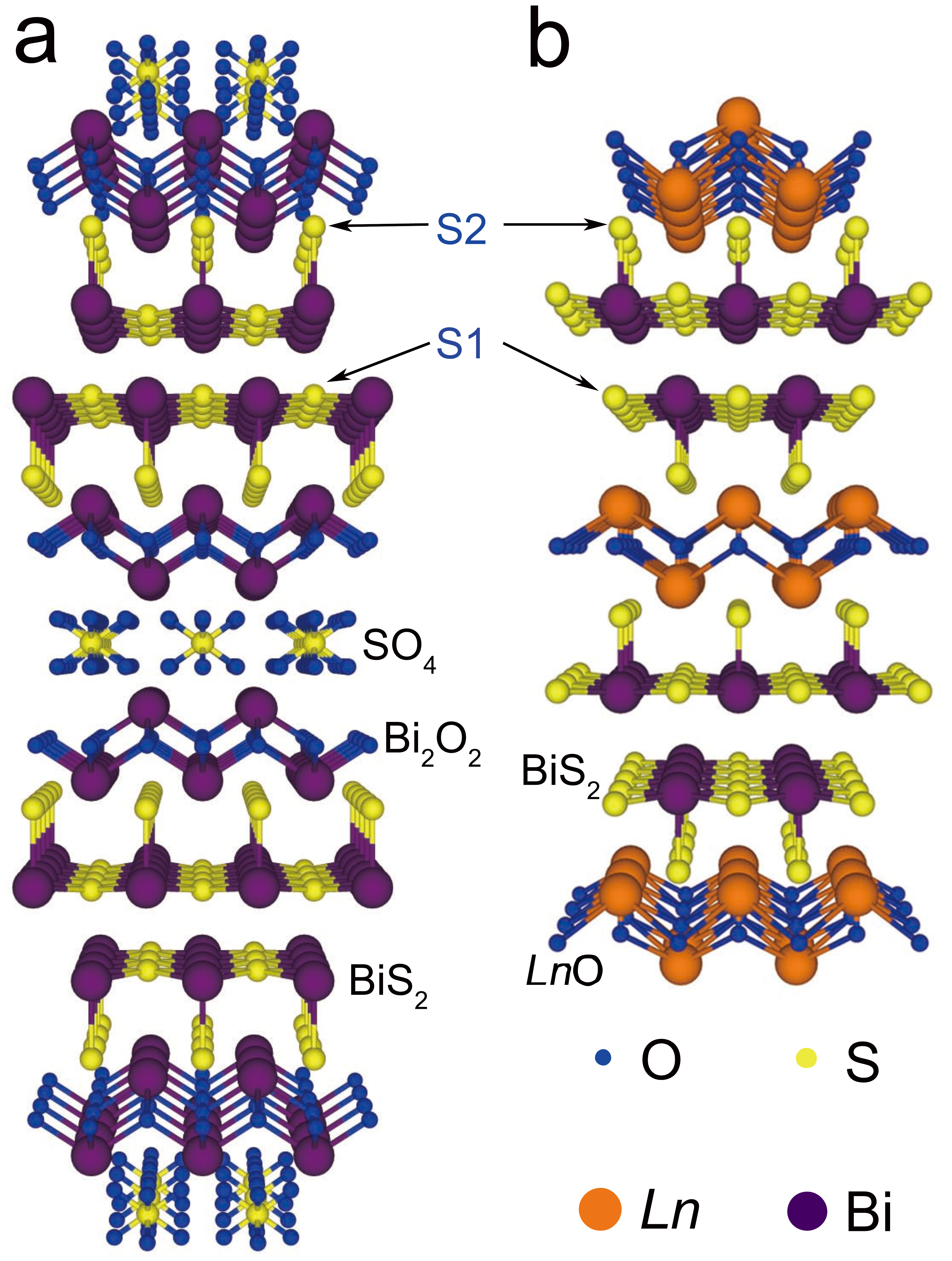}
\caption{(Color online) Representative crystal structure of (a) Bi$_{4}$O$_{4}$S$_{3}$ and (b) \textit{Ln}OBiS$_{2}$. O$^{2-}$ ions in the \textit{Ln}O blocking layer can be partially replaced by F$^{-}$. S1 and S2 indicate the in-plane and out-of-plane positions of S, respectively, in the superconducting BiS$_{2}$ layers of the two structures.}
\label{FIG.1.}
\end{figure}

Currently, superconductivity has been induced in five different parent compounds \textit{Ln}(O,F)BiS$_{2}$ (\textit{Ln} = La, Ce, Pr, Nd, Yb), with the substitution of F for O \cite{yazici2013superconductivity}. Polycrystalline samples of these compounds can be synthesized by using solid state reactions in sealed quartz tubes. However, along with the main superconducting phase, small amounts of secondary phases including \textit{Ln}F$_{3}$ and BiF$_{3}$ may also be formed in the polycrystalline samples \cite{mizuguchi2012superconductivity,fang2015enhancement,yazici2013superconductivity,jeon2014effect,demura2013new,xing2012superconductivity}. As a result, the actual chemical composition of these compounds may be different from the nominal chemical composition, particularly regarding the concentration of F, resulting in an overestimated level of electron doping into the BiS$_2$ layers. This concern has also been raised in the study of single-crystalline samples of NdO$_{1-x}$F$_{x}$BiS$_2$, in which case, the actual electron doping level was reported to be much lower than anticipated \cite{ye2014electronic}. Hence, $T_c$ values of samples with the same nominal chemical composition reported by different authors may differ from one another \cite{yazici2013superconductivity,mizuguchi2014superconductivity}. Although conclusive evidence of inhomogeneity in polycrystalline samples of the BiS$_{2}$-based superconductors has not been reported, measurements of electrical resistivity ($\rho$), magnetic susceptibility ($\chi$), and specific heat ($C$), etc. may still be partially affected by possible inhomogeneity of the samples, and thus compromise reports of some of the intrinsic properties of these compounds. In addition, the existence of defects and disorder in these materials may also complicate the analysis of experiments and comparison with theoretical models. As mentioned by Kuroki, establishing the intrinsic phase diagram of the BiS$_2$-based superconductors is still a significant challenge in research on BiS$_{2}$-based superconductors \cite{kuroki2014what}.

\subsection{Electronic and crystal structures, and their correlation with $T_c$}

\begin{figure}[h]
\centering
\includegraphics[width=8cm]{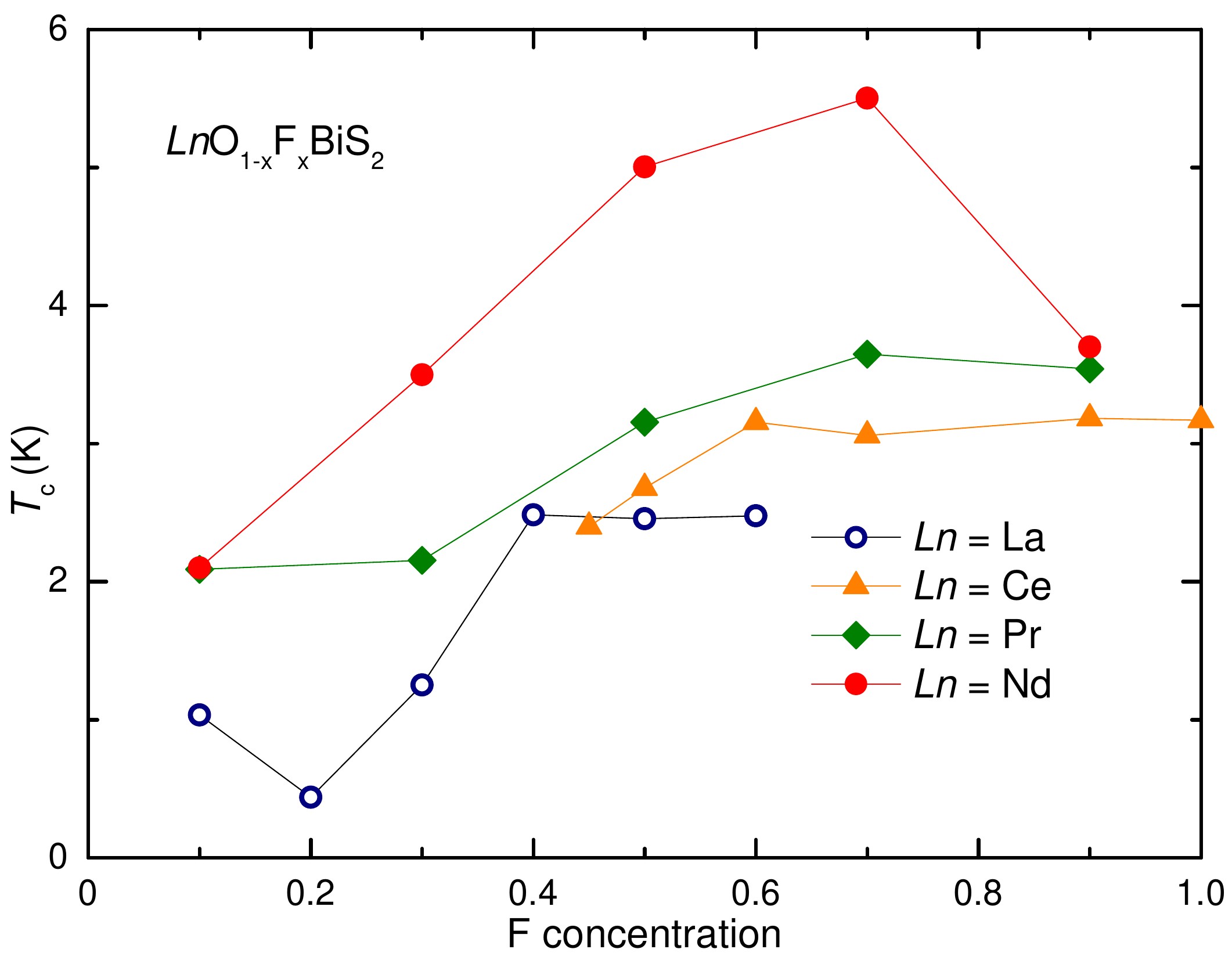}
\caption{(Color online) Dependence of $T_c$ on nominal F concentration for  \textit{Ln}O$_{1-x}$F$_{x}$BiS$_{2}$ (\textit{Ln} = La-Nd). The $T_c$ values for \textit{Ln} = La, Ce, Pr, Nd  are obtained from Ref. \cite{higashinaka2014low,demura2015coexistence,jha2014superconducting,jha2014superconducting2}, respectively. For interpretation of the definition of $T_c$ of each sample, the reader is referred to the corresponding articles.}
\label{FIG.2.}
\end{figure}

The $T_c$ vs F concentration plots for \textit{Ln}O$_{1-x}$F$_{x}$BiS$_{2}$, displayed in Fig. 2, show a broad range of F concentration in which superconductivity can be induced; there is also an apparent enhancement of $T_c$ with increasing \textit{Ln} atomic number. The optimal F concentration for superconductivity in these compounds falls within the range of 0.4 $\leqslant$ $x$ $\leqslant$ 0.7. As the concentration of F is increased, the superconductivity of these compounds was also found to be more resistant to the destructive effects of an external magnetic field up to the F solubility limit \cite{jha2014superconducting2,jha2014superconducting}. As an example, the upper critical field of NdO$_{1-x}$F$_{x}$BiS$_{2}$ increases from 0.9 T at $x$ = 0.3 to 3.3 T at $x$ = 0.7 \cite{jha2014superconducting2}. For LaO$_{1-x}$F$_{x}$BiS$_{2}$, first-principles calculations suggest that the density of states at the Fermi level increases with increasing F substitution and attains its maximum value at $x$ = 0.5 \cite{yildirim2013ferroelectric}. Hence, a large number of studies have been made on compounds with a nominal composition given by \textit{Ln}O$_{0.5}$F$_{0.5}$BiS$_{2}$ \cite{xing2012superconductivity,demura2013new,mizuguchi2012superconductivity}.

Electronic structure calculations for \textit{Ln}O$_{1-x}$F$_{x}$BiS$_{2}$, and, in particular, for the compound LaO$_{0.5}$F$_{0.5}$BiS$_{2}$, have been widely reported \cite{morice2013electronic,usui2012minimal}. It was suggested that LaOBiS$_{2}$ is an insulator with an energy gap of $\sim$0.8 eV \cite{li2013phonon}, which is consistent with experimental evidence revealing significantly high values of $\rho$ and negative temperature coefficients ($d$$\rho$/$dT$ $\textless$ 0) in polycrystalline samples \cite{lee2013crystal}. As F is substituted into the O site (see Fig. 1),  it is expected that electrons will be doped into the conduction bands which consist mainly of Bi-6p orbitals, resulting in an increase in charge carrier density in the BiS$_{2}$ layers \cite{usui2012minimal,mizuguchi2012bis2}. The possibility of conventional strong coupling $s$-wave superconductivity for Bi$_4$O$_4$S$_3$, LaO$_{0.5}$F$_{0.5}$BiS$_{2}$, and  NdO$_{1-x}$F$_x$BiS$_2$ has been emphasized by magnetic penetration depth measurements by means of muon-spin spectroscopy or tunnel diode resonator techniques \cite{shruti2013evidence,lamura2013s,jiao2015evidence,biswas2013low}, and some theoretical studies also suggest that BiS$_2$-based compounds are strong electron-phonon coupled superconductors \cite{yildirim2013ferroelectric,wan2013electron,li2013phonon}. However, other unconventional pairing mechanisms including extend $s$- or $d$-wave pairing, $g$-wave pairing, and triplet spin pairing were also proposed \cite{martins2013RPA,yang2013triplet,liang2014pairing,wu2014g,zhou2013probing}.  While evidence for both electron-phonon coupling \cite{shruti2013evidence,lamura2013s,jiao2015evidence,biswas2013low,yildirim2013ferroelectric,wan2013electron,li2013phonon} and strong electron-electron interactions \cite{martins2013RPA,yang2013triplet,liang2014pairing,zhou2013probing,wu2014g,li2013strong,liu2014giant} has been reported, the concentration of charge carriers in the BiS$_2$ layers is considered to be essential for the superconductivity of the \textit{Ln}O$_{1-x}$F$_{x}$BiS$_{2}$ compounds  \cite{mizuguchi2012superconductivity,yazici2013superconductivity,kotegawa2012pressure,deguchi2013evolution,kajitani2014correlation,lamura2013s,lee2013crystal,li2013phonon,awana2013appearance}.

\begin{figure}[t]
\centering
\includegraphics[width=8cm]{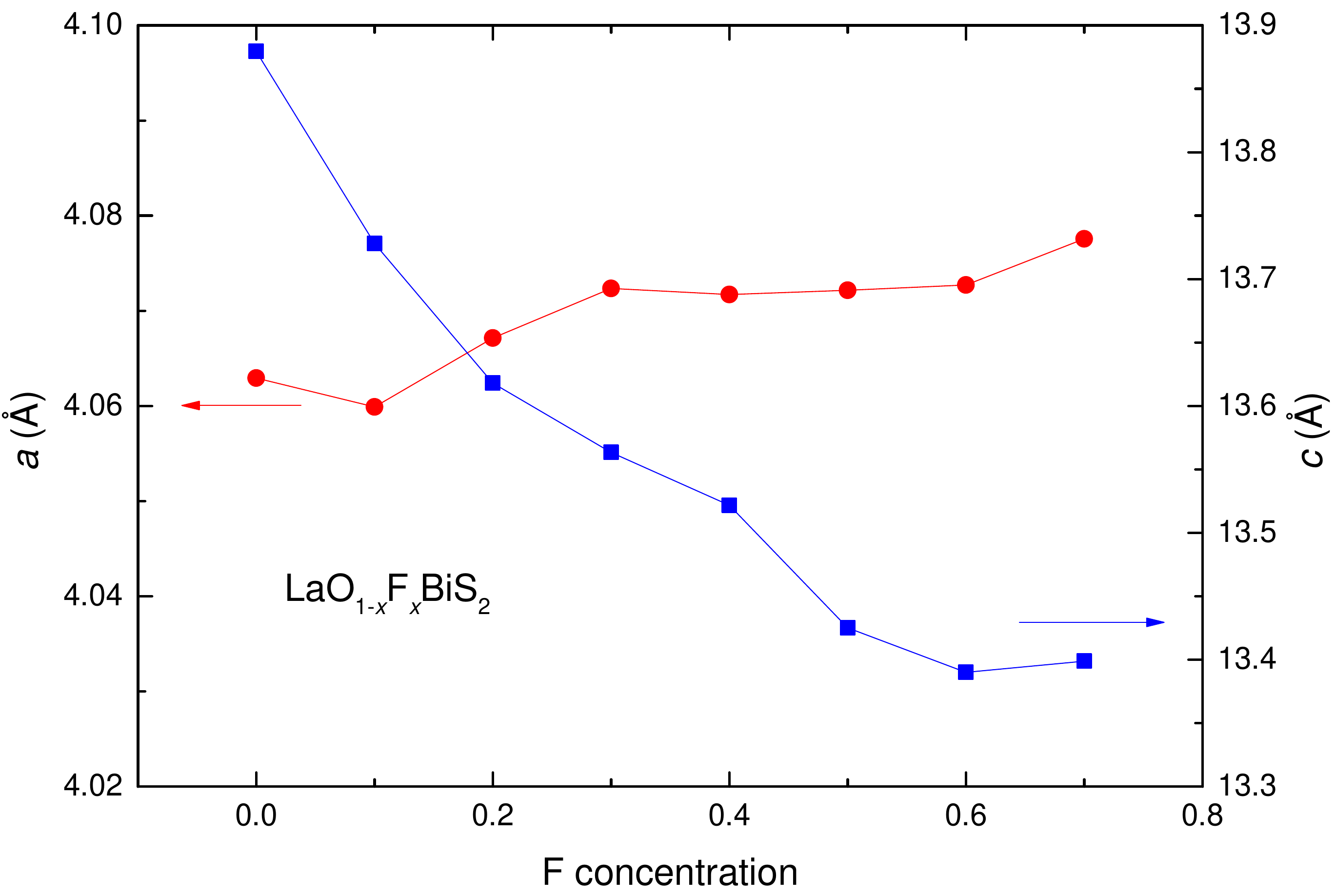}
\caption{(Color online) Nominal F concentration dependence of the lattice constants $a$ (red) and $c$ (blue) of AG LaO$_{1-x}$F$_{x}$BiS$_{2}$. The data are obtained from Ref. \cite{deguchi2013evolution}.}
\label{FIG.3.}
\end{figure}

\begin{figure}[t]
\centering
\includegraphics[width=7cm]{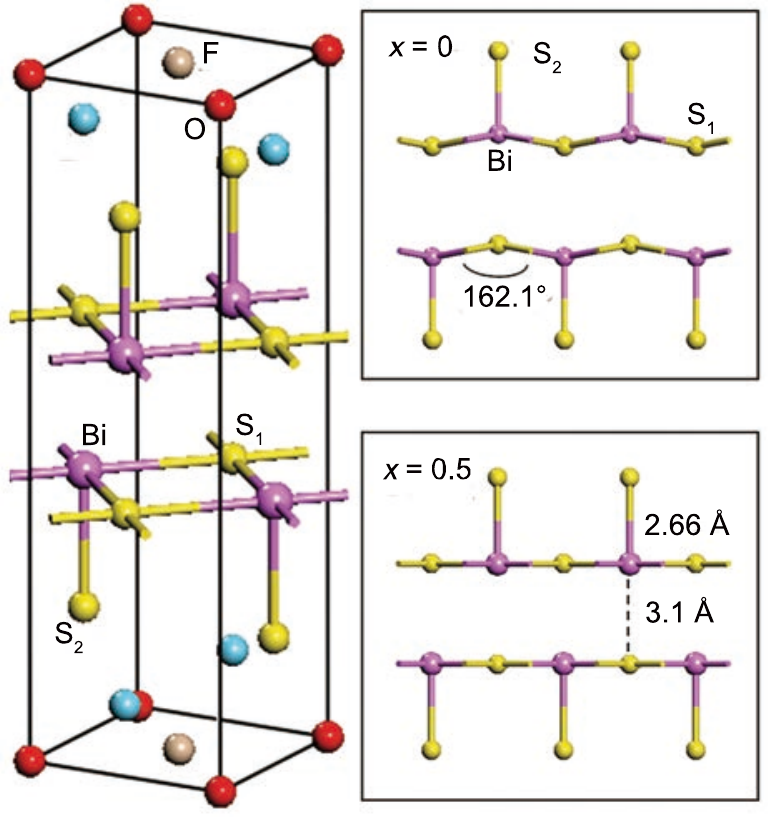}
\caption{(Color online) (left) Crystal structure of LaO$_{0.5}$F$_{0.5}$BiS$_{2}$. (right) The side view of the BiS$_{2}$ layers for $x$ = 0 and $x$ = 0.5 is displayed in the top and bottom panel, respectively. Adapted with permission from Ref. \cite{yildirim2013ferroelectric}. Copyrighted by the American Physical Society.}
\label{FIG.4.}
\end{figure}

The substitution of F for O in \textit{Ln}OBiS$_{2}$ may also induce a change in lattice parameters due to the smaller size of the F$^{-}$ ions compared with O$^{2-}$ ions. For \textit{Ln}O$_{1-x}$F$_{x}$BiS$_{2}$, the length of the $a$-axis does not change significantly upon F substitution; however, the $c$-axis decreases monotonically until the F concentration reaches the solubility limit (typically around 50\%) as illustrated by the example of LaO$_{1-x}$F$_{x}$BiS$_{2}$ shown in  Fig.3 \cite{mizuguchi2012superconductivity,jha2014superconducting,xing2012superconductivity,jha2014superconducting2,demura2013new}. The highest value of $T_c$ was reported when $c$ attains its minimum value, and the evolution of $T_c$ seems consistent with the change in lattice parameter $c$ \cite{mizuguchi2012superconductivity,jha2014superconducting,jha2014superconducting2}. However, it should be noted that these results do not necessarily indicate an exclusive role of the lattice parameter $c$ in affecting $T_c$, since it is observed that the level of electron doping also reaches a maximum when the $c$ parameter is smallest.

\begin{figure}[!t]
\centering
\includegraphics[width=8cm]{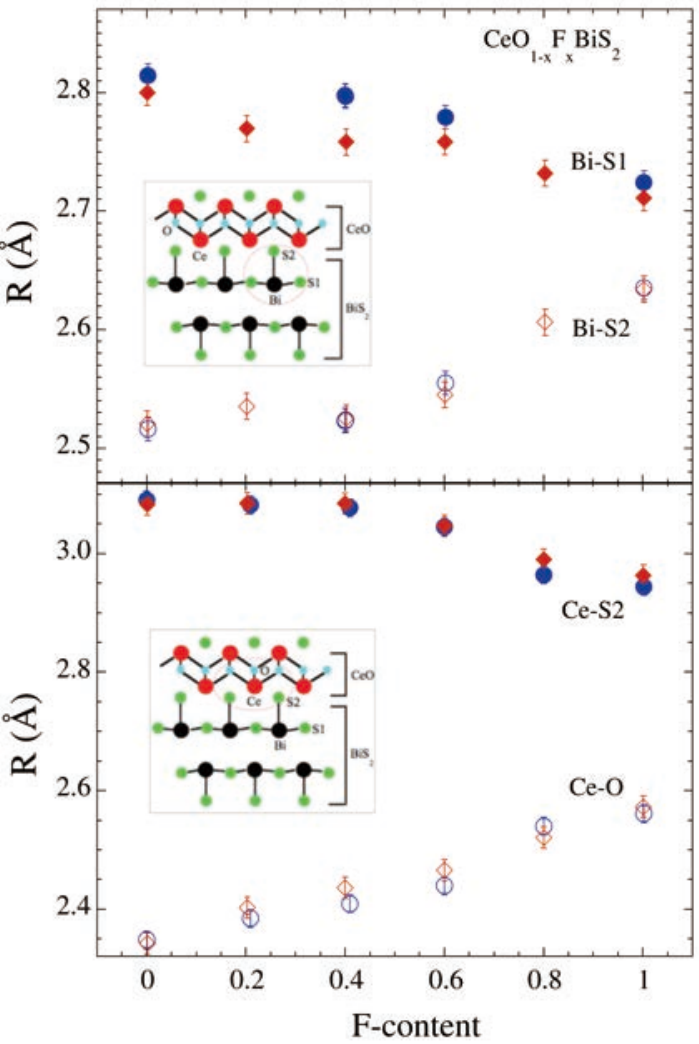}
\caption{(Color online)  Near-neighbor distances from Bi (upper) and Ce (lower) as a function of F concentration. Blue circles and red diamonds represent the data taken from the as-grown and high pressure annealed samples, respectively. The insets show structural cartoons with near-neighbour atomic clusters (encircled) around Bi and Ce.  Reprinted with permission from Ref. \cite{paris2014determination}: Determination of local atomic displacements in CeO$_{1-x}$F$_x$BiS$_2$ system, J. Phys.: Condens. Matter 26 (2014) 435701. Copyright (2014) by the IOP Publishing.}
\label{FIG.5.}
\end{figure}

Local distortion of the lattice may also be induced by F substitution. Asymmetric broadening of the diffraction peaks with increasing F concentration in LaO$_{1-x}$F$_{x}$BiS$_{2}$ was observed in neutron scattering measurements, suggesting that strain may be induced along the $c$-axis due to a random substitution of the F ions at O sites \cite{lee2013crystal}. It is also possible for a distortion to occur in the BiS$_{2}$ plane as a result of F substitution. For the LaO$_{1-x}$F$_{x}$BiS$_{2}$ compounds, first-principles calculations show that the $z$ coordinate of the in-plane S and Bi atoms in the unit cell decreases with doping, yielding a nearly perfect planar structure at $x$ = 0.5 (see Fig. 4). In fact, by using single crystal x-ray diffraction, nearly flat in-plane S-Bi-S angles were observed in LaO$_{1-x}$F$_{x}$BiS$_{2}$  at $x$ = 0.46, which is close to the optimal F concentration value ($x$ = 0.5) that yields the highest value of $T_c$  \cite{miura2014crystal, miura2014structure}. A flat Bi–S plane would result in an increase in hybridization of the $p$$_{x}$/$p$$_{y}$ orbitals of Bi and S and, in turn, enhanced superconductivity. Investigation of extended x-ray absorption fine structure (EXAFS) of the CeO$_{1-x}$F$_{x}$BiS$_{2}$ compounds reveals that the local structure of both the BiS$_{2}$ superconducting layer and the Ce(O,F) blocking layer changes systematically with increasing F concentration \cite{paris2014determination}. The Bi-S2 distance increases with F concentration while the Ce-S2 distance decreases with F concentration (see Fig. 5), leading to the breaking of the Ce-S-Bi coupling channel \cite{sugimoto2014role}. Consequently, the hybridization between the Ce 4$f$ orbital and the Bi 6$p$ conduction band decreases with F concentration and the Ce$^{3+}$/Ce$^{4+}$ valence fluctuations are suppressed, resulting in a completely localized 4$f$$^{1}$ (Ce$^{3+}$) state at $x$ $\textgreater$ 0.4, where the samples exhibit superconductivity and ferromagnetism \cite{sugimoto2014role}. Therefore, it has been suggested that the valence of Ce, i.e., the coupling between the Ce 4$f$ and Bi 6$p$ states, is not beneficial to the superconductivity.

\begin{figure}[!t]
\centering
\includegraphics[width=8cm]{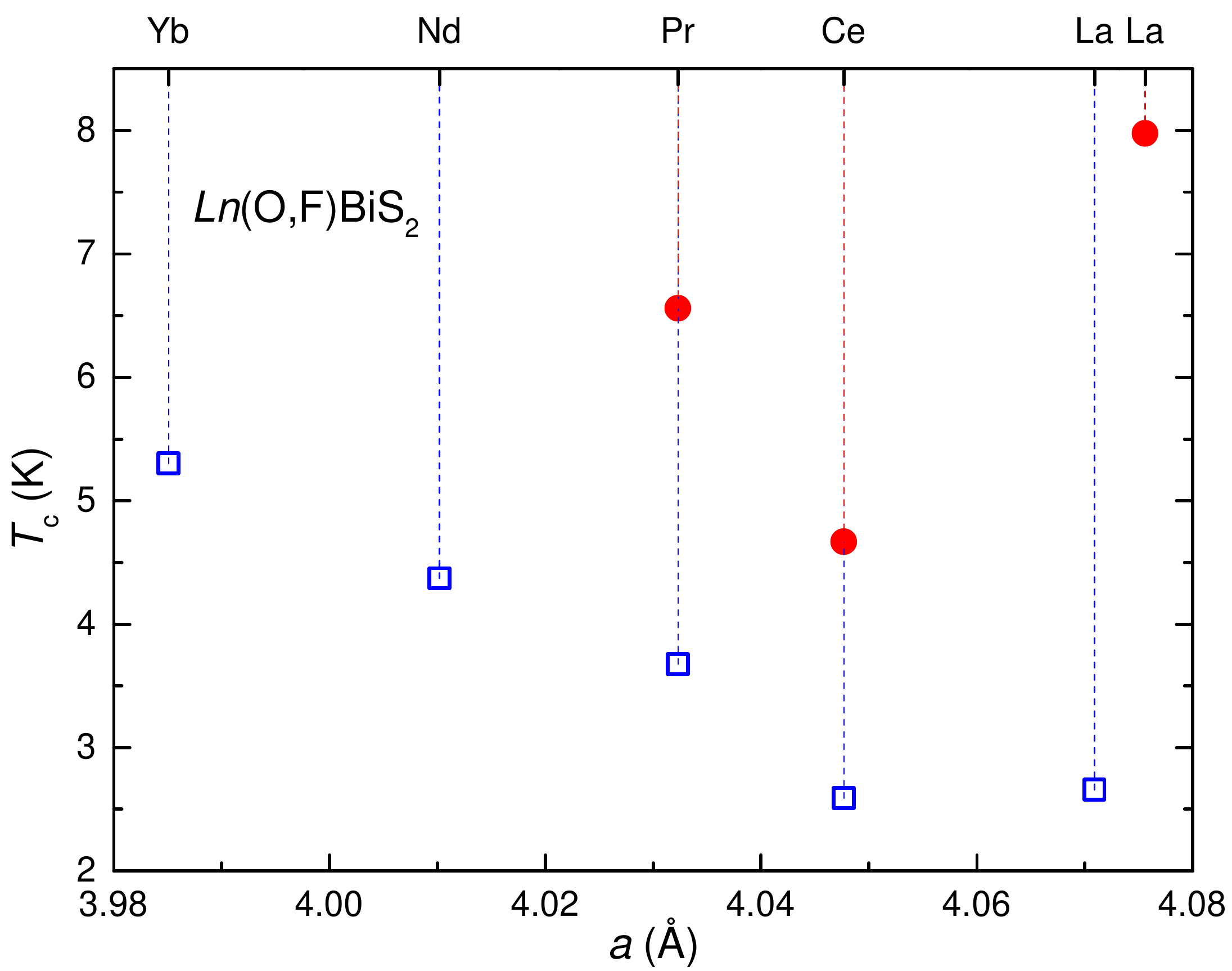}
\caption{(Color online) Correlation between $T_c$ and $a$-axis of \textit{Ln}O$_{1-x}$F$_{x}$BiS$_{2}$ ( $x$ = 0.5 for \textit{Ln} = La, Pr, Nd, Yb, $x$ = 0.3 for \textit{Ln} = Ce) superconductors. The open squares and filled circles represent the data obtained from as-grown samples and high pressure annealed samples, respectively. Dashed lines are guides to the eye. The $T_c$ and $a$ values for \textit{Ln} = Nd, Yb samples are obtained from Ref. \cite{yazici2013superconductivity} and the data for \textit{Ln} = La, Ce, Pr samples are taken from Refs. \cite{kajitani2014correlation,kajitani2015increase,kajitani2014enhancement}, respectively.}
\label{FIG.6.}
\end{figure}

Considering the contraction in radius for the \textit{Ln} ions, one may expect changes in both lattice parameters $a$ and $c$ as well as changes in $T_c$ with increasing atomic number of \textit{Ln}. Figure 6 displays typical values of $T_c$, defined as the temperature at which $\rho$ falls to 50\% of its normal-state value, for the compounds in the \textit{Ln}(O,F)BiS$_{2}$ system as a function of lattice constant $a$. Except for the \textit{Ln} =  Yb and Nd cases, the $T_c$ values of the samples after HP annealing are higher than the values of the AG samples. In addition, the dependence of $T_c$ on the lattice paramter $a$ for the AG samples are completely different from the HP annealed samples: for the AG samples, $T_c$ increases with decreasing $a$ (heavier \textit{Ln}); however, the value of $T_c$ for LaO$_{0.5}$F$_{0.5}$BiS$_{2}$ after HP annealing (a larger value of $a$) is significantly higher than the value of other HP annealed samples (smaller values of $a$). While the values of $T_c$ for both AG and HP annealed samples appear to be correlated, although differently, with the length of the $a$-axis \cite{mizuguchi2014review}, no structural phase transition was observed by HP annealing and the lattice parameters $a$ and $c$ for the AG and HP annealed samples are similar; this suggests that there should be other factors which have a significant effect on $T_c$. The markedly different values in $T_c$ between the AG and HP annealed samples currently has no satisfactory explanation. However, evidence of a local distortion in the lattice has been reported, and this distortion is likely to be a source of the difference in $T_c$. For the HP annealed samples of CeO$_{1-x}$F$_{x}$BiS$_2$, the in-plane Bi-S1 distance is shorter; however, the out-of-plane Bi-S2 distance and the sizing of the blocking layers appear to be marginally effected by the HP annealing (see Fig. 5) \cite{paris2014determination}. It was also suggested that the crystal structure within the $ab$ plane of the CeO$_{0.3}$F$_{0.7}$BiS$_{2}$ compound evolved to a higher symmetry after HP annealing, resulting in a better tetragonal phase \cite{kajitani2015increase}. Finally, the uniaxial strain along the $c$-axis, which was found in annealed samples of the LaO$_{0.5}$F$_{0.5}$BiS$_{2}$ and PrO$_{0.5}$F$_{0.5}$BiS$_{2}$ compounds, was reported to be positively correlated with the enhancement of superconductivity \cite{kajitani2014correlation,kajitani2014enhancement}. These results indicate that the optimization of the local structure is important for realizing higher $T_c$ values in the BiS$_2$-based superconductors.

\subsection{Normal state electrical resistivity, magnetic susceptibility, and specific heat}

\begin{figure}[h]
\centering
\includegraphics[width=8cm]{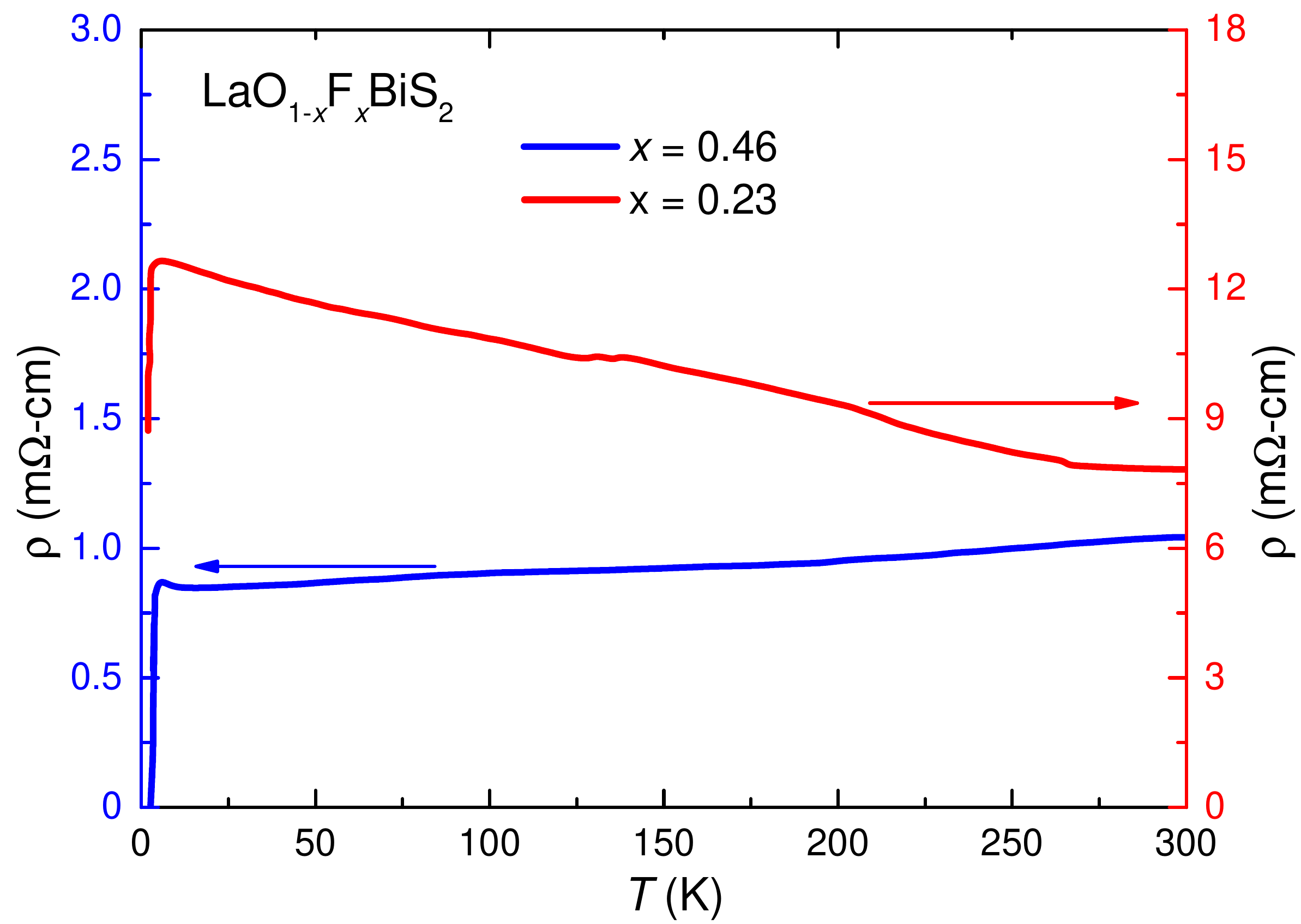}
\caption{(Color online) Temperature dependence of the electrical resistivity for LaO$_{1-x}$F$_{x}$BiS$_{2}$ ($x$ = 0.23 and 0.46) single crystals. The data are obtained from Ref. \cite{miura2014crystal}.}
\label{FIG.7.}
\end{figure}

Given that F substitution enhances the charge carrier density of BiS$_2$-based parent compounds, one would expect that F-substitution may induce more metallic-like behavior in these materials. The temperature dependence of $\rho$($T$) in the normal state for single crystals of LaO$_{1-x}$F$_{x}$BiS$_{2}$ ($x$ = 0.23, 0.46) is shown in Fig. 7; while the crystal with $x$ = 0.23 exhibits semiconducting-like behavior ($d$$\rho$/$dT$ $\textless$ 0), the crystal with $x$ = 0.46 shows metallic behavior ($d$$\rho$/$dT$ $\textgreater$ 0) with significantly reduced values of resistivity \cite{miura2014crystal}. These results are consistent with electronic band structure calculations which indicate that the parent compounds in this system should behave as band insulators but become metallic upon doping electron carriers into the BiS$_2$ layers \cite{usui2012minimal}. However, the theoretical calculations do not fit well with measurements of $\rho$($T$) of polycrystalline samples of \textit{Ln}OBiS$_{2}$ which tend to exhibit strong sample dependent results. Some of the \textit{Ln}OBiS$_{2}$ parent compounds, which are in polycrystalline form,  have been reported to show semimetallic behavior in contrast to the semiconducting-like behavior of the corresponding F-doped polycrystalline samples \cite{yazici2013superconductivity2,jha2014superconducting,jha2014superconducting2,demura2015coexistence,xing2012superconductivity}. This behavior may be related to the quality of these polycrystalline samples, since complete semiconducting-like behavior is also reported in the case of LaOBiS$_{2}$ and in the case of  polycrystalline CeOBiS$_{2}$, the resistivity can be significantly increased together with the appearance of semiconducting-like behavior after HP annealing \cite{lee2013crystal,demura2015coexistence}.

Substitution of F into the lattice does not always result in a suppression of semiconducting-like behavior in  polycrystalline samples. In fact, F substitution can enhance semiconducting-like behavior which has been observed in AG samples of CeO$_{1-x}$F$_{x}$BiS$_{2}$ and HP annealed samples of LaO$_{1-x}$F$_{x}$BiS$_{2}$ \cite{deguchi2013evolution,xing2012superconductivity}; this is in contrast to observations of suppressed semiconducting-like behavior by F substitution in single-crystalline samples of CeO$_{1-x}$F$_{x}$BiS$_{2}$ and LaO$_{1-x}$F$_{x}$BiS$_{2}$ \cite{miura2014crystal,nagao2014growth}. Although the cause of such unusual behavior is still unclear, the existence of grain boundaries as well as poor intergrain contacts in polycrystalline samples may contribute significantly to the semiconducting-like behavior of the F-substituted compounds. In addition to the semiconducting-metallic transition induced by F substitution in single crystals of LaO$_{1-x}$F$_{x}$BiS$_{2}$, research on single crystals of NdO$_{1-x}$F$_{x}$BiS$_{2}$ reveals metallic behavior in the normal state along the $ab$-plane of the crystal; this is different from the semiconducting behavior observed in polycrystalline samples of NdO$_{1-x}$F$_{x}$BiS$_{2}$ \cite{jha2014superconducting2,nagao2013structural,liu2014giant,ye2014electronic,nagao2014growth}. However, reports of the electrical transport properties in the normal state of single crystals of CeO$_{1-x}$F$_{x}$BiS$_{2}$ reveal semiconducting-like behavior in the $ab$-plane which is consistent with measurements on polycrystalline samples of the same compound \cite{miura2014structure}. Hence, the intrinsic electrical transport properties of the \textit{Ln}O$_{1-x}$F$_{x}$BiS$_{2}$ compounds seem to be  dependent on a combination of lanthanide element \textit{Ln}, F concentration $x$, and quality of the samples.

Both zero-field-cooled (ZFC) and field-cooled (FC) measurements of $\chi$($T$) were performed for \textit{Ln}O$_{1-x}$F$_x$BiS$_2$ samples with different \textit{Ln} and F concentrations. As shown in Fig. 8, ZFC measurements on \textit{Ln}O$_{0.5}$F$_{0.5}$BiS$_{2}$ (\textit{Ln} = La, Pr, Nd) yield a significant shielding volume fraction with $T_c$ onset values that are consistent with the $\rho$($T$) data, while FC measurements  reveal that  $\chi$($T$) remains nearly unchanged in the superconducting state relative to the normal state, indicating strong vortex pinning \cite{yazici2013superconductivity}. For the case of samples with \textit{Ln} = Ce and Yb and nominal F concentrations of $x$ $\geqslant$ 0.7 and $x$ = 0.5, respectively, magnetic phase transitions were also observed, apart from diamagnetic signals arising from the appearance of superconductivity \cite{yazici2013superconductivity,kajitani2015increase,nagao2014growth}. In addition to the induction of superconductivity and enhancement of $T_c$ as a result of F substitution, the shielding volume fraction may be significantly enhanced at those temperatures where it has begun to saturate as a result of the increase in F concentration; a typical example of this behavior was found for NdO$_{1-x}$F$_{x}$BiS$_{2}$ reported in Ref. \cite{jha2014superconducting}.

\begin{figure}[t]
\centering
\includegraphics[width=8cm]{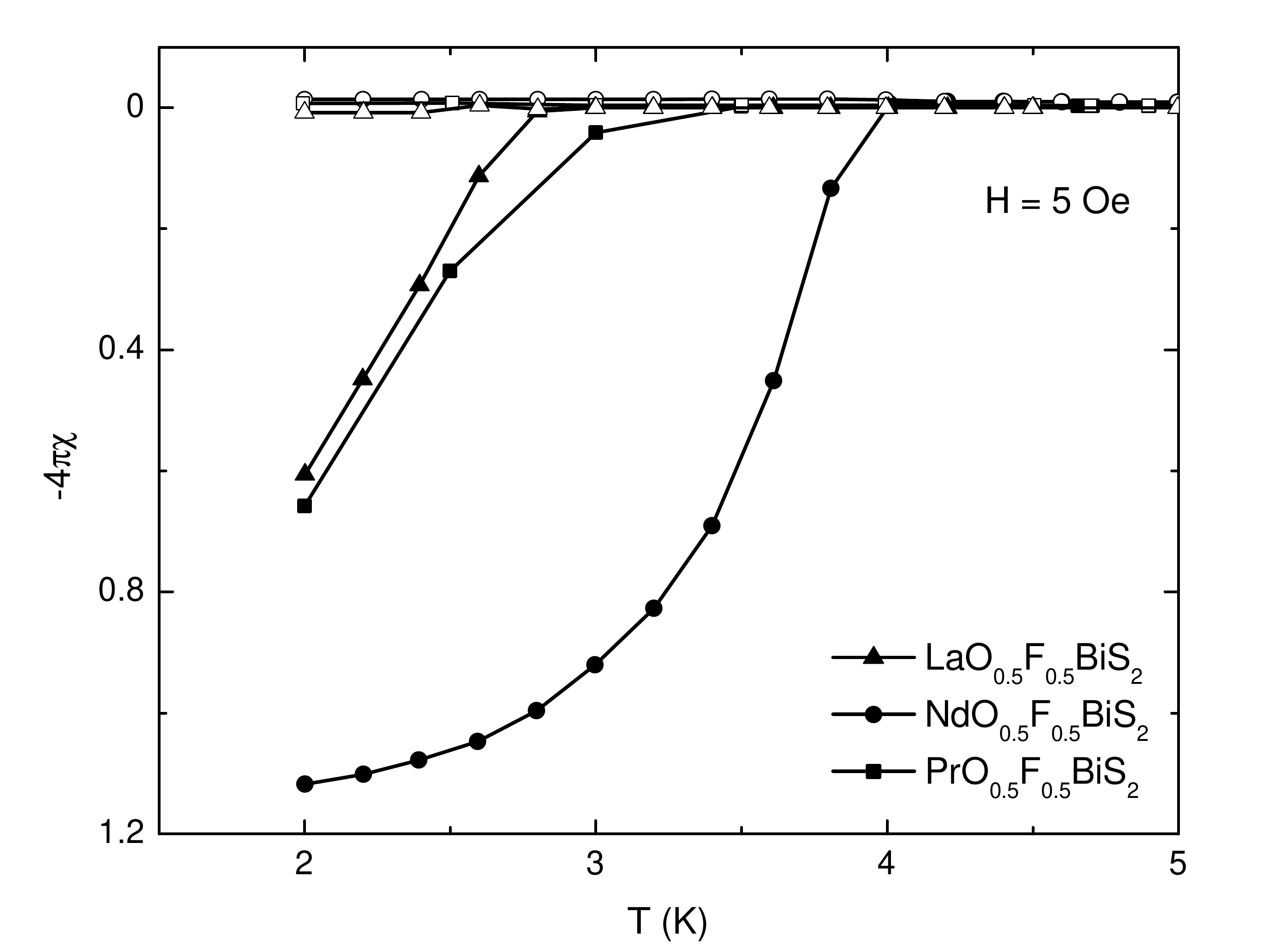}
\caption{Temperature dependence of the magnetic susceptibility $\chi$  for \textit{Ln}O$_{0.5}$F$_{0.5}$BiS$_{2}$ (\textit{Ln} = La, Pr, Nd) measured in the field-cooled (FC) (open symbols) and zero-field-cooled (ZFC) (filled symbols) processes. The data are taken from Ref. \cite{yazici2013superconductivity}.}
\label{FIG.8.}
\end{figure}

Figure 9 displays specific heat $C$($T$) data for the polycrystalline LaO$_{0.5}$F$_{0.5}$BiS$_2$ compound in the temperature range 1.8 K $\leqslant$ $T$ $\leqslant$ 50 K \cite{yazici2013superconductivity}.  The expression $C(T)/T = \gamma + \beta T^{2}$, where  $\gamma$ is the electronic specific heat coefficient and  $\beta$ is the coefficient of the lattice contribution, was used in to fit the $C/T$ vs $T^2$ data in the normal state which is displayed in the inset in the lower right hand side of Fig. 9. Shown in the inset in the upper left hand side of Fig. 9 are $C_e$/$T$ vs $T$ data, where $C_e$ is the electronic contribution to the specific heat. Bulk superconductivity in polycyrstalline LaO$_{0.5}$F$_{0.5}$BiS$_{2}$ was confirmed with the appearance of a clear jump in the $C_e$/$T$ vs $T$ data at $T_c$. The ratio of the observed jump to the electronic contribution to the specific heat at $T_c$ is estimated to be $\Updelta$$C$/$\upgamma$$T_{c}$ = 0.94, which is lower than the weak coupling BCS value of 1.43, but large enough to confirm bulk superconductivity in this compound. For the other \textit{Ln}O$_{0.5}$F$_{0.5}$BiS$_{2}$ compounds with \textit{Ln} = Ce, Pr, Nd, and Yb, the significant upturns in the $C$/$T$ vs $T$ data observed at low temperature are likely due to the presence of a Schottky-like anomaly for \textit{Ln} = Pr and Nd or magnetic ordering for \textit{Ln} = Ce and Yb \cite{yazici2013superconductivity,yazici2015superconductivity}. No clear $C$($T$)/$T$ jump related to the appearance of superconductivity was observed in these samples \cite{yazici2013superconductivity}. The jumps in $C$($T$)/$T$ in these compounds may not be readily observable because they may be reduced by the pairbreaking interaction due to the magnetic ions; the jumps may also be obscured by the broadening of the superconducting transitions by sample inhomogeneity or by the large upturns in $C$($T$)/$T$. More rigorous experimental and theoretical studies on higher quality samples are needed to establish the nature of superconductivity in these compounds.

\begin{figure}[t]
\centering
\includegraphics[width=8cm]{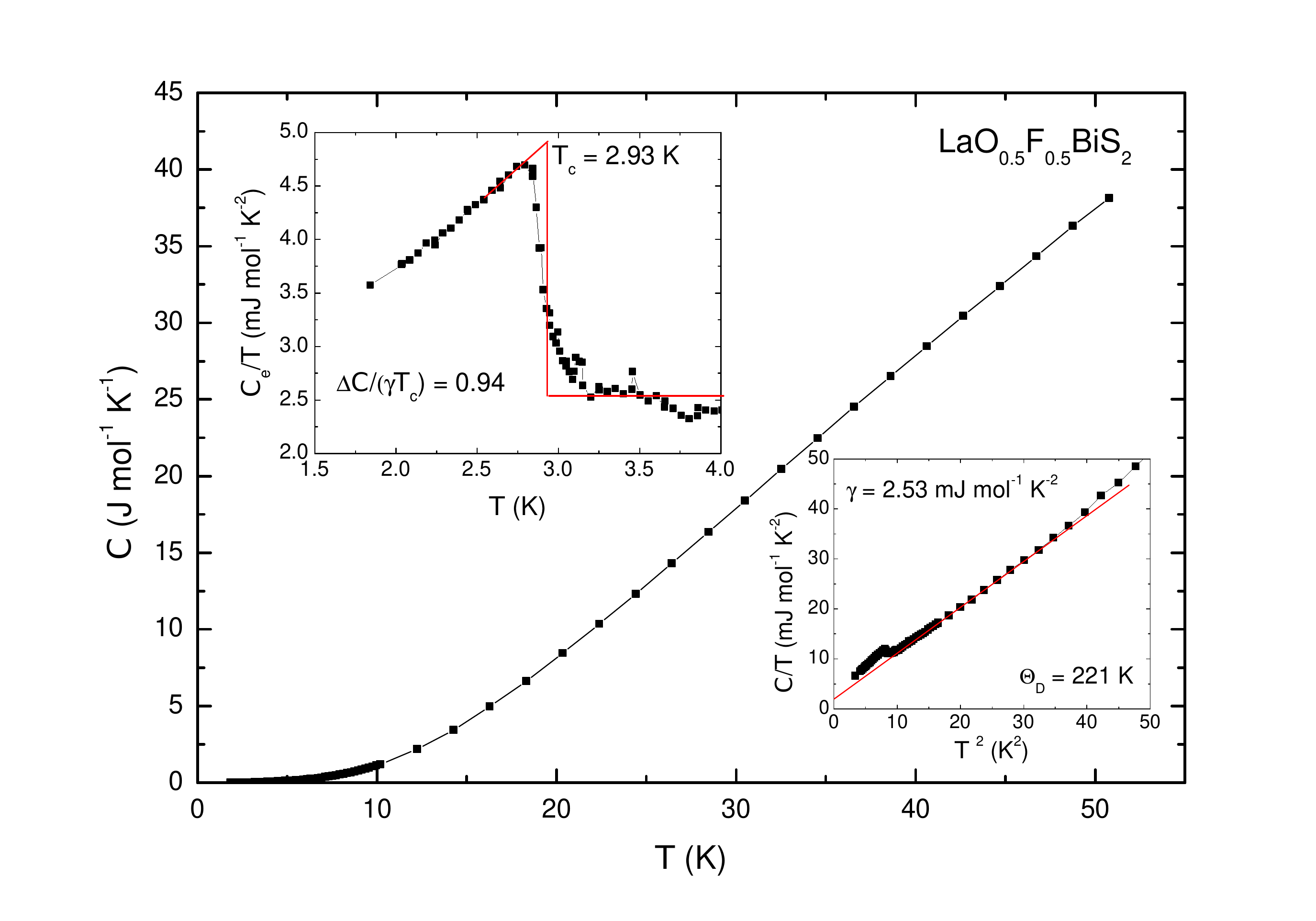}
\caption{(Color online) Specific heat $C$ vs temperature $T$ for LaO$_{0.5}$F$_{0.5}$BiS$_{2}$. A plot of $C/T$ vs $T$$^{2}$ is shown in the inset in the lower right hand part of the figure. The red line is a fit of the expression $C$($T$)/$T$ = $\gamma$ + $\beta$$T$$^{2}$. A plot of $C$$_{e}$/$T$ vs $T$, where $C$$_{e}$ is the electronic contribution to the specific heat, in the vicinity of the superconducting transition is shown in the inset in the upper left hand side of the figure. The data are taken from Ref. \cite{yazici2013superconductivity}.}
\label{FIG.9.}
\end{figure}

\section{La$_{1-x}$$M$$_x$OBiS$_2$ ($M$ = Th, Hf, Zr, and Ti)}

\begin{figure}[h]
\centering
\includegraphics[width=8cm]{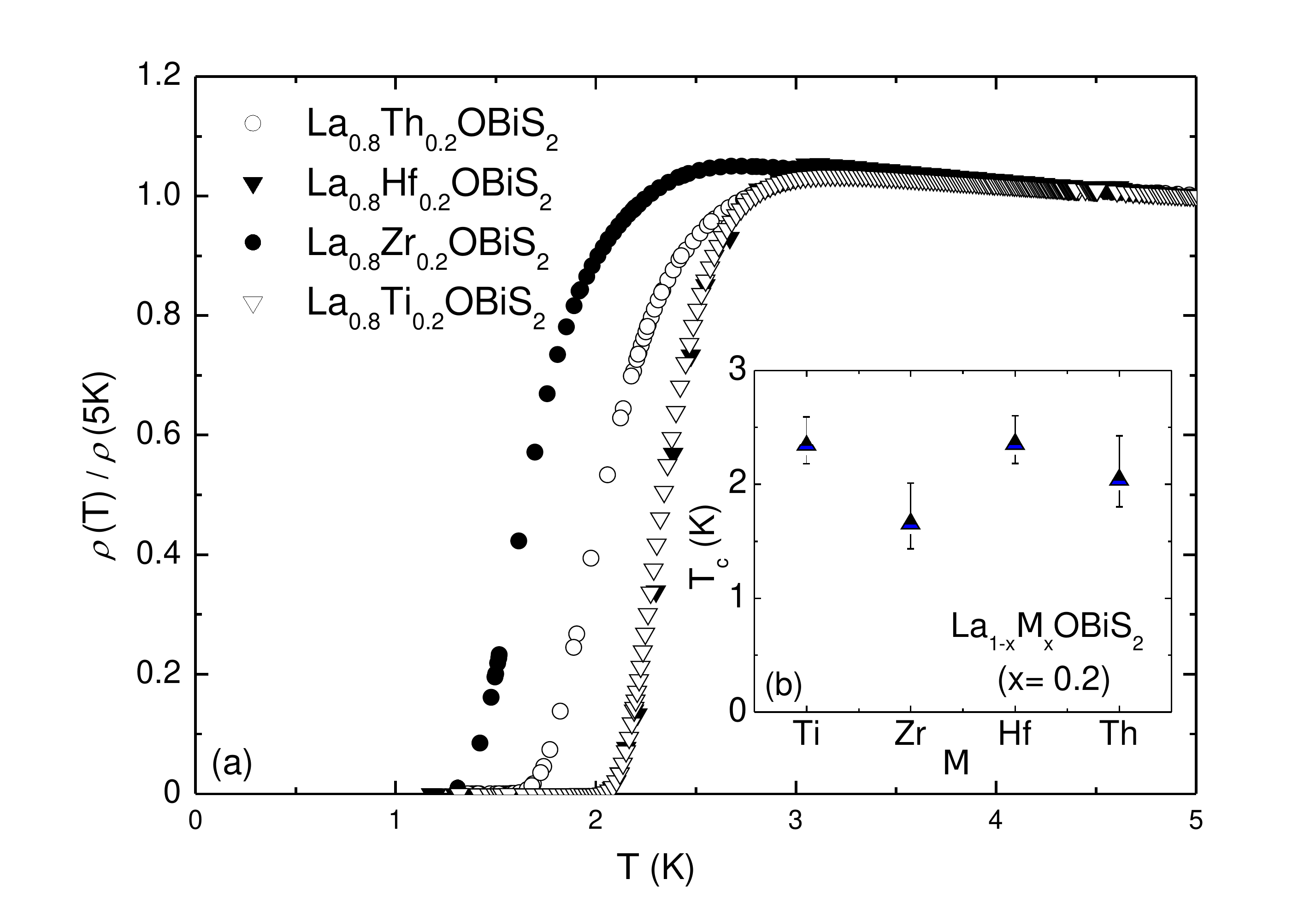}
  \caption{Resistive superconducting transitions and (inset) superconducting transition temperatures $T_c$ for La$_{1-x}$$M_x$OBiS$_2$ compounds with $M$ = Th, Hf, Zr, Ti and $x = 0.2$. $T_c$ is defined as the temperature where $\rho$ drops to 50$\%$ of its normal state value, and the width of the transition is determined from the temperatures where the resistivity drops to 90$\%$ and 10$\%$ of its normal state value. Reprinted with permission from Ref. \cite{yazici2013superconductivity2}. Copyright (2013) by the American Physical Society.}
\label{FIG.10.}
\end{figure}

As an alternative to F substitution, electron doping in the LaOBiS$_2$ system is possible by partially replacing trivalent La with tetravalent $M$ ($M$ = Th, Hf, Zr, and Ti) \cite{yazici2013superconductivity2}. It has been reported that the La$_{1-x}$$M$$_x$OBiS$_2$ system exhibits superconductivity with values of $T_c$ up to 2.85 K. A comparison of $T_c$ values observed in the various compounds is displayed in the inset of Fig. 10 in which 20\% ($x$ = 0.2) of $M$ = Th, Hf, Zr, and Ti was substituted for La in La$_{1-x}$$M$$_x$OBiS$_2$; there appears to be no clear correlation between $T_c$ and the atomic radius of the tetravalent atom that replaces the La ion. The lowest value of $T_c$ is observed for the La$_{0.8}$Zr$_{0.2}$OBiS$_{2}$ compound; while sharper superconducting transitions are observed for the La$_{0.8}$Hf$_{0.2}$OBiS$_2$ and La$_{0.8}$Ti$_{0.2}$OBiS$_2$ compounds, broader transitions are observed for the La$_{0.8}$Zr$_{0.2}$OBiS$_2$ and La$_{0.8}$Th$_{0.2}$OBiS$_2$ compounds. X-ray diffraction experiments show that both the $a$ and $c$ lattice parameters do not change significantly with different $M$$^{4+}$ substitutions at the $x$ = 0.2 concentration. However, with increasing $M$ concentration, both values of $a$ and $c$ for La$_{1-x}$$M$$_x$OBiS$_2$ were observed to decrease gradually. For La$_{1-x}$Th$_x$OBiS$_2$, $T_c$ first decreases with $x$ from 2.85 K at $x$ = 0.15 to 2.05 K at $x$ = 0.20 and then remains roughly constant at higher concentration. A similar evolution of $T_c$ with increasing $M$ was also observed for La$_{1-x}$$M$$_x$OBiS$_2$ ($M$ = Hf, Ti); however, after partially replacing La$^{3+}$ with Sr$^{2+}$, no superconductivity was observed in La$_{1-x}$Sr$_x$OBiS$_2$ down to $\sim$1 K in the range $0.1 \leqslant x \leqslant 0.3$, which is in agreement with the behavior reported for La$_{1-x}$Mg$_{x}$OBiS$_2$ ($x$ = 0-0.2) samples \cite{yazici2013superconductivity2,chen2013effect}. These results suggest that hole doping is not sufficient to induce superconductivity and reveals the importance of electron doping in achieving superconductivity in BiS$_{2}$-based compounds.

\begin{figure}[t]
\centering
\includegraphics[width=8cm]{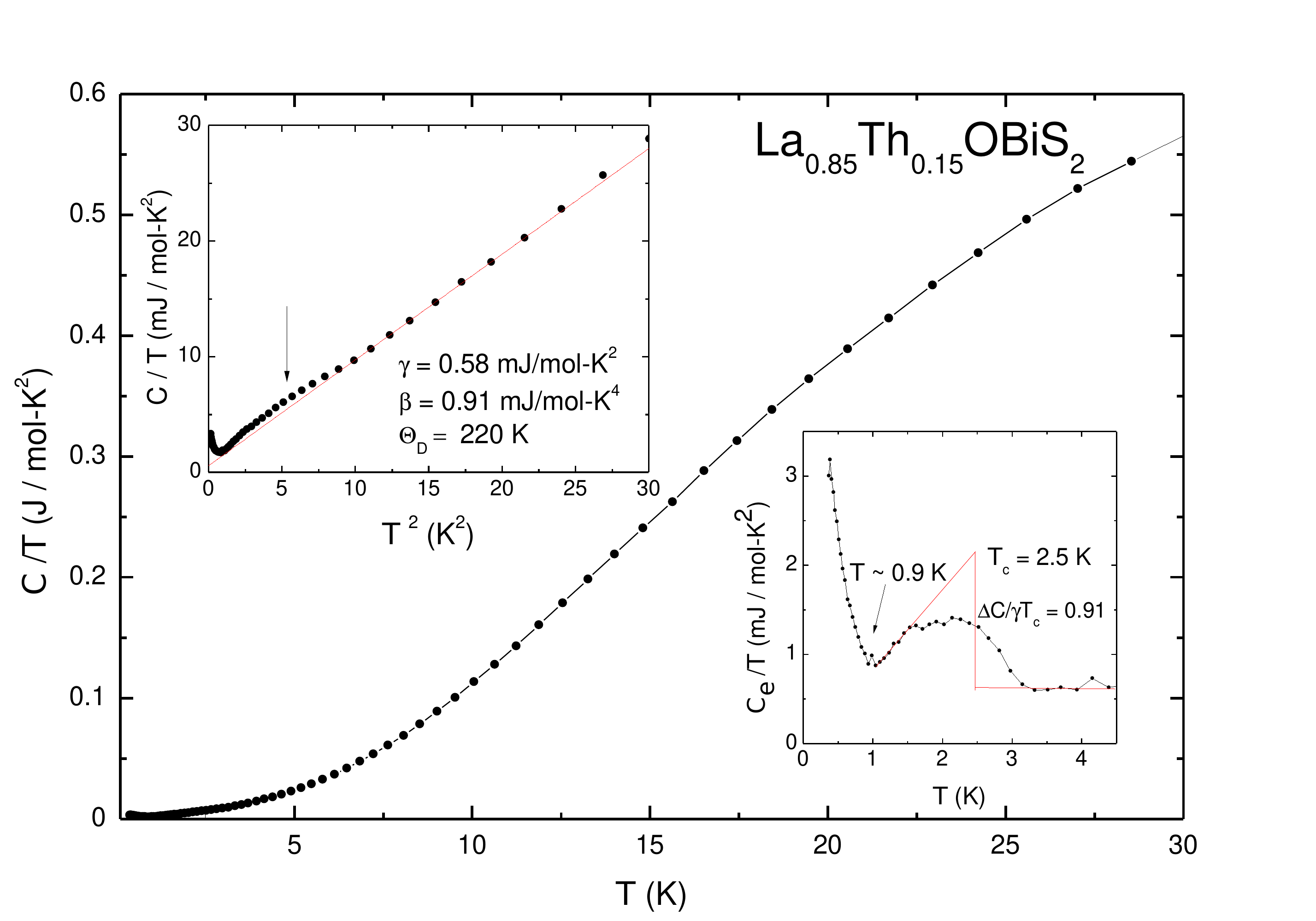}
  \caption{(Color online) Specific heat $C$ divided by temperature $T$, $C / T$,  vs $T$ for La$_{0.85}$Th$_{0.15}$OBiS$_2$.  $C / T$ vs $T^2$ is shown in the inset in the upper left hand part of the figure.  The red line is a fit of the expression $C$($T$)/$T$ = $\gamma$ + $\beta$$T^{2}$ to the data. The inset in the lower right part of the figure shows a plot of $C_e$($T$) vs $T$ in the vicinity of the superconducting transition. Reprinted with permission from Ref. \cite{yazici2013superconductivity2}. Copyright (2013) by the American Physical Society.}

\label{FIG.11.}
\end{figure}

Measurements of $\rho$($T$) for the La$_{0.85}$Th$_{0.15}$OBiS$_2$ and La$_{0.8}$Hf$_{0.2}$OBiS$_2$ compounds were performed in various magnetic fields down to 0.36 K \cite{yazici2013superconductivity2}. As magnetic field is increased, the superconducting  transition width broadens and the onset of superconductivity gradually shifts to lower temperatures. The orbital critical fields $H_{c2}(0)$ for the La$_{0.85}$Th$_{0.15}$OBiS$_2$ and La$_{0.8}$Hf$_{0.2}$OBiS$_2$ compounds were inferred from their initial slopes of $H_c$ with respect to $T$ using the conventional one-band WHH theory \cite{WHH}, yielding values of 1.09 and 1.12 T, respectively. These values of $H_{c2}(0)$ are close to the values of 1.9 T for $H_{c2}(0)$ observed in LaO$_{0.5}$F$_{0.5}$BiS$_2$ \cite{yazici2013superconductivity, awana2013appearance}, suggesting that there is a common superconducting phase characteristic of BiS$_2$-based superconductors.

Specific heat measurements were performed for the La$_{0.85}$Th$_{0.15}$OBiS$_2$ and La$_{0.8}$Hf$_{0.2}$OBiS$_2$ compounds in the temperature range of 0.36 K $\leqslant T \leqslant$ 30 K \cite{yazici2013superconductivity2}. A clear feature is observed between $1$ and $3$ K for La$_{0.85}$Th$_{0.15}$OBiS$_2$ as shown in the inset at the lower right hand side of Fig. 11. A value of $T_c$ = 2.5 K was estimated from an idealized entropy conserving construction, which is close to the $T_c$ obtained from $\rho$($T$) measurement ($T_c$ = 2.85 K). The presence of the feature suggests that superconductivity is a bulk phenomenon in this compound.  The value of $\Delta C / \gamma T_c$ for La$_{0.85}$Th$_{0.15}$OBiS$_2$ is estimated to be $\sim$0.91, which is less than the value of 1.43 predicted by the BCS theory, but similar to that seen in LaO$_{0.5}$F$_{0.5}$BiS$_2$ \cite{yazici2013superconductivity}. However, only a small feature in $C_e / T$ data for La$_{0.8}$Hf$_{0.2}$OBiS$_2$ is observed around the $T_c$ (2.36 K) obtained from $\rho$($T$) measurements. The absence of a well-defined superconducting jump at $T_c$ for La$_{0.8}$Hf$_{0.2}$OBiS$_2$ is likely to be a consequence of the superconducting transition being spread out in temperature due to sample inhomogeneity \cite{yazici2013superconductivity}.



\section{Chemical substitution effects on \textit{Ln}(O,F)BiS$_{2}$}

It has been demonstrated that the charge carrier (electron) density is important for the superconductivity of BiS$_{2}$-based compounds. In addition,  values of $T_c$ for AG samples of \textit{Ln}O$_{0.5}$F$_{0.5}$BiS$_{2}$ increase with increasing atomic number of the \textit{Ln} component as shown in Fig. 6 \cite{mizuguchi2014superconductivity,yazici2013superconductivity}. (The ionic size of the \textit{Ln$^{3+}$} ion decreases with increasing atomic number.) This size contraction has the effect of reducing the dimensions of the blocking layers which contain the \textit{Ln$^{3+}$} ions, thereby squeezing the neighboring  BiS$_2$ layers, i.e., effectively generating chemical pressure on the superconducting BiS$_2$ layers. As discussed in section 3.1, it appears that there is a correlation between $T_c$ and the decrease in the size of the $a$-axis in the \textit{Ln}(O,F)BiS$_{2}$ system.  Hence, substitution of the element La with heavier lanthanide elements such as Sm, Eu, and Gd holds promise as a strategy for further enhancement of $T_c$ in the BiS$_2$-based superconducting compounds. Recently, the compound with nominal chemical composition  SmO$_{0.5}$F$_{0.5}$BiS$_{2}$ was synthesized and reported to be nonsuperconducting \cite{thakur2015synthesis}; attempts have been unsuccessful in synthesizing other \textit{Ln}O$_{0.5}$F$_{0.5}$BiS$_{2}$ compounds with \textit{Ln} = Eu-Tm. However, there has been recent success in synthesizing compounds in which there is a partial substitution at the \textit{Ln} site with relatively smaller atoms as a method for enhancing $T_c$ and further determining which parameters are relevant in affecting superconductivity \cite{kajitani2014chemical,fang2015enhancement,thakur2015synthesis,kajitani2015chemical,chen2013effect,mizuguchi2015in,jeon2014effect}.

\begin{figure}[t]
\centering
\includegraphics[width=8cm]{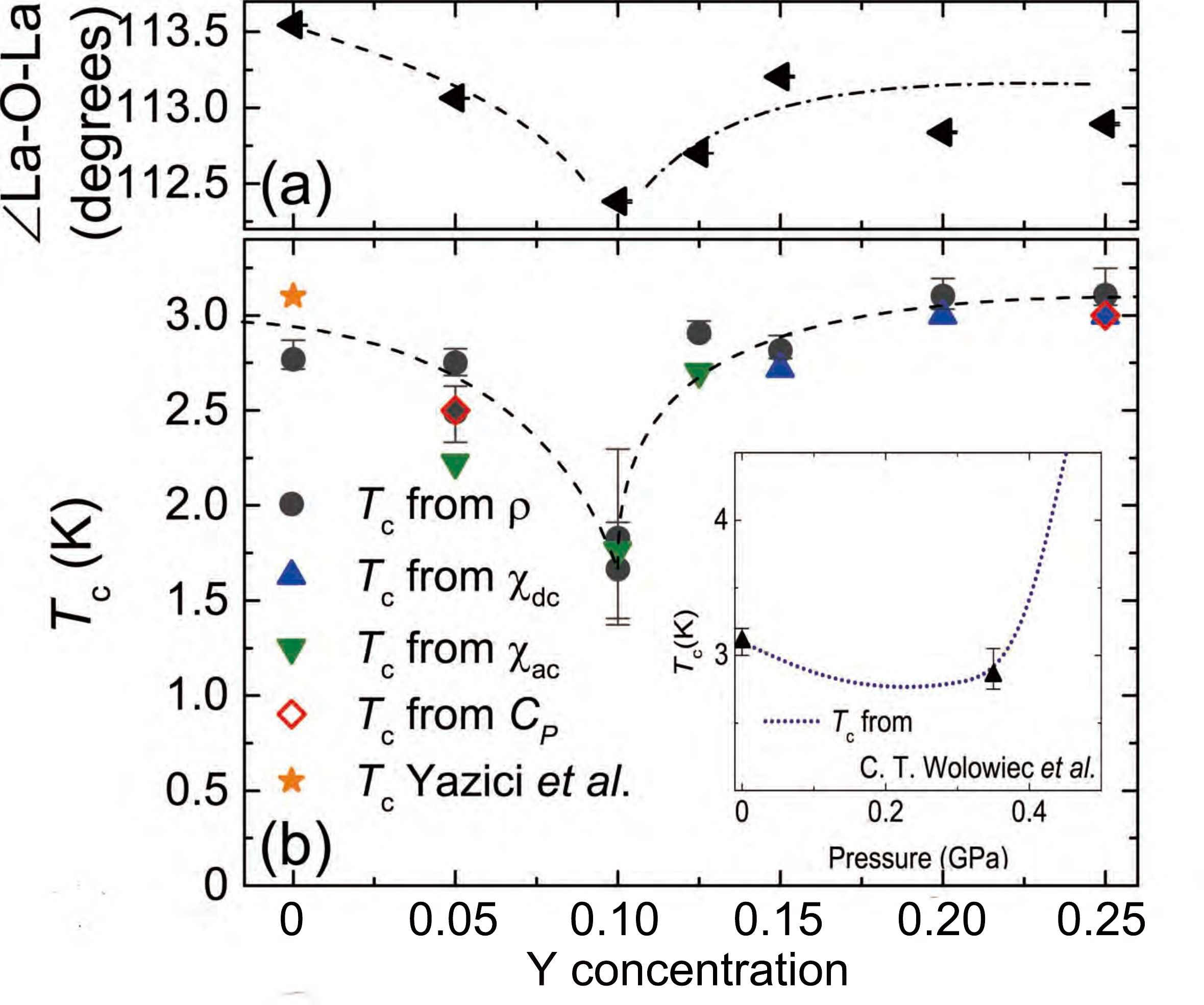}
  \caption{(Color online) (a) La(Y)-O(F)-La(Y) bond angle vs nominal yttrium concentration. (b) Dependence of $T_c$ on Y concentration  for the La$_{1-x}$Y$_{x}$O$_{0.5}$F$_{0.5}$BiS$_{2}$ system. The orange star is the $T_c$ of LaO$_{0.5}$F$_{0.5}$BiS$_{2}$ reported in Ref. \cite{yazici2013superconductivity}. The inset shows the behavior of $T_c$ under hydrostatic pressure \cite{wolowiec2013pressure}. Reprinted with permission from Ref. \cite{jeon2014effect}. Copyright (2014) by the American Physical Society.}
\label{FIG.12.}
\end{figure}


As an example, partial substitution of La with the smaller Y ions in the La$_{1-x}$Y$_{x}$O$_{0.5}$F$_{0.5}$BiS$_{2}$ compound allows one to study of the effects of chemical pressure on superconductivity \cite{jeon2014effect}. Similar to La, Y has a trivalent electronic configuration and no magnetic moment, making Y$^{3+}$ a suitable replacement for the La$^{3+}$ ions. As a result of the substitution of Y for La in  La$_{1-x}$Y$_{x}$O$_{0.5}$F$_{0.5}$BiS$_{2}$, $T_c$ was found to gradually decrease from 2.8 K at $x$ = 0 to 1.8 K at $x$ = 0.1 and then remain roughly constant at $\sim$3.0 K for $x$ $\geqslant$ 0.125 as shown in Fig. 12. The lattice parameter $a$ and the unit cell volume of La$_{1-x}$Y$_{x}$O$_{0.5}$F$_{0.5}$BiS$_{2}$ monotonically decrease with increasing Y concentration until the solubility limit $x$ = 0.20. Interestingly, the evolution of the La-O-La bond angle with increasing $x$ closely follows the evolution of $T_c$($x$) as shown in Fig. 12. It should be mentioned that the lattice constant $c$ changes in the opposite manner (decreasing with increasing Y concentration), which suggests that $T_c$ may also be affected by the value of $c$.  Unfortunately, due to the solubility limit of Y in La$_{1-x}$Y$_{x}$O$_{0.5}$F$_{0.5}$BiS$_{2}$, further investigation of the relation between lattice parameters, bond angle, and $T_c$ by the method of substitution is not possible and the highest chemical pressure obtained in La$_{0.8}$Y$_{0.2}$O$_{0.5}$F$_{0.5}$BiS$_{2}$ is still lower than 0.7 GPa, above which a structural phase transition from tetragonal ($P4/nmm$) to monoclinic ($P21/m$) has been observed in LaO$_{0.5}$F$_{0.5}$BiS$_2$ under applied external hydrostatic pressure \cite{wolowiec2013pressure,tomita2014pressure}.


\begin{figure}[t]
\centering
\includegraphics[width=8cm]{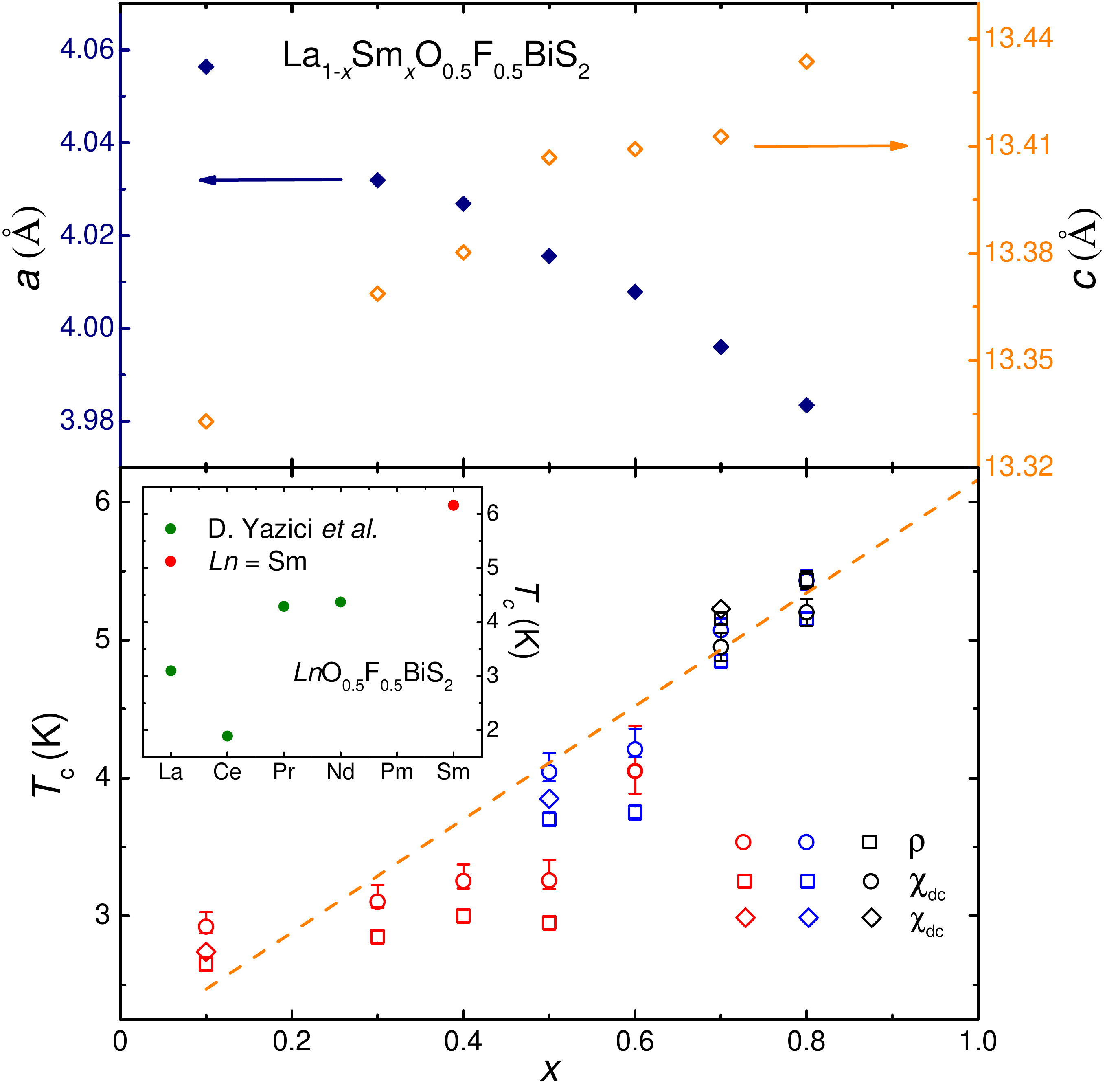}
  \caption{(Color online) Dependence of the lattice constants $a$ and $c$ (upper panel) and $T_c$ (lower panel) on Sm concentration $x$ of La$_{1-x}$Sm$_{x}$O$_{0.5}$F$_{0.5}$BiS$_{2}$. Red, blue, and purple symbols in the lower panel represent results for samples annealed at 800\celsius, 750\celsius, and 710\celsius, respectively. The inset shows representative values of $T_c$ of \textit{Ln}(O,F)BiS$_{2}$ reported in Ref. \cite{yazici2013superconductivity}. The other data are obtained from Ref. \cite{fang2015enhancement}.}
\label{FIG.13.}
\end{figure}

Figure 13 shows the Sm concentration dependence of $T_c$ for La$_{1-x}$Sm$_{x}$O$_{0.5}$F$_{0.5}$BiS$_{2}$. An enhancement of superconductivity is observed in the La$_{1-x}$Sm$_{x}$O$_{0.5}$F$_{0.5}$BiS$_{2}$ compound in which $T_c$ increases from $\sim$2.8 K at $x$ = 0.1 to 5.4 K at the solubility limit of $x$ = 0.8 \cite{fang2015enhancement}. A linear extrapolation of the $T_c$($x$) data beyond the $x$ = 0.8 solubility limit to $x$ = 1.0 allows for an estimate of the value of $T_c$ for the parent compound SmO$_{0.5}$F$_{0.5}$BiS$_{2}$ to be as high as $\sim$6.2 K; this value of $T_c$ is significantly higher than values of $T_c$ reported for other AG samples of the \textit{Ln}O$_{0.5}$F$_{0.5}$BiS$_{2}$ compound and is therefore consistent with the trend that higher values of $T_c$ are observed in \textit{Ln}O$_{0.5}$F$_{0.5}$BiS$_{2}$ compounds containing a \textit{Ln} element having a larger atomic number (see the inset of Fig. 13). As the Sm concentration increases from $x$ = 0.1 to 0.8, the lattice constant $c$ increases while the lattice constant $a$ gradually decreases, resulting in an overall suppression of the unit cell volume of $\sim$3\%, however, no structural phase transition is observed. In comparison, there is only a $\sim$2\% decrease in the unit cell volume of LaO$_{0.5}$F$_{0.5}$BiS$_{2}$ before the pressure-induced structure phase transition \cite{tomita2014pressure}. Hence, the effects of Sm substitution for La and applied hydrostatic pressure on the compound LaO$_{0.5}$F$_{0.5}$BiS$_{2}$ seem somewhat different.

Effects of substitutions at the \textit{Ln} site in the \textit{Ln}O$_0.5$F$_0.5$BiS$_2$ compounds were also investigated at various F concentrations which allows for the study of the tuning of the lattice constant $c$ and its effects on superconductivity. Phase diagrams of $T_c$ vs \textit{Ln} substitution for the Ce$_{1-x}$Nd$_{x}$O$_{1-y}$F$_{y}$BiS$_{2}$ and Nd$_{1-z}$Sm$_{z}$O$_{1-y}$F$_{y}$BiS$_{2}$ systems were determined from measurements of magnetization at nominal F concentrations of $y$ = 0.7, 0.5, and 0.3 as shown in Fig. 14 \cite{kajitani2015chemical}. Superconductivity is induced and noticeably enhanced for those compounds containing higher concentrations of \textit{Ln} elements with relatively smaller atomic size (or larger atomic number), i.e., higher concentrations of Nd and Sm atoms. The shift in the phase boundary to the right for decreasing F concentration suggests the importance of F concentration in the induction of superconductivity in these BiS$_2$-based compounds. For a given F concentration, it appears that superconductivity is enhanced in compounds with $Ln$ ions of smaller ionic size; however, there is no appreciable increase in $T_c$ for greater concentrations of F. It is interesting to note that samples with different nominal values of F concentration that share similar values of $T_c$ also exhibit similar values of the ratio of the length of the $c$ axis to that of the $a$ axis (see Fig. 14(b)); this implies that the superconductivity observed in the BiS$_2$-based compounds is closely related with the lattice constant ratio $c$/$a$. The effect of charge carrier density on the enhancement of superconductivity in these BiS$_2$ compounds seems to be of less importance in comparison to the lattice parameter $a$ and the $c$/$a$ ratio.

\begin{figure}[t]
\centering
\includegraphics[width=8cm]{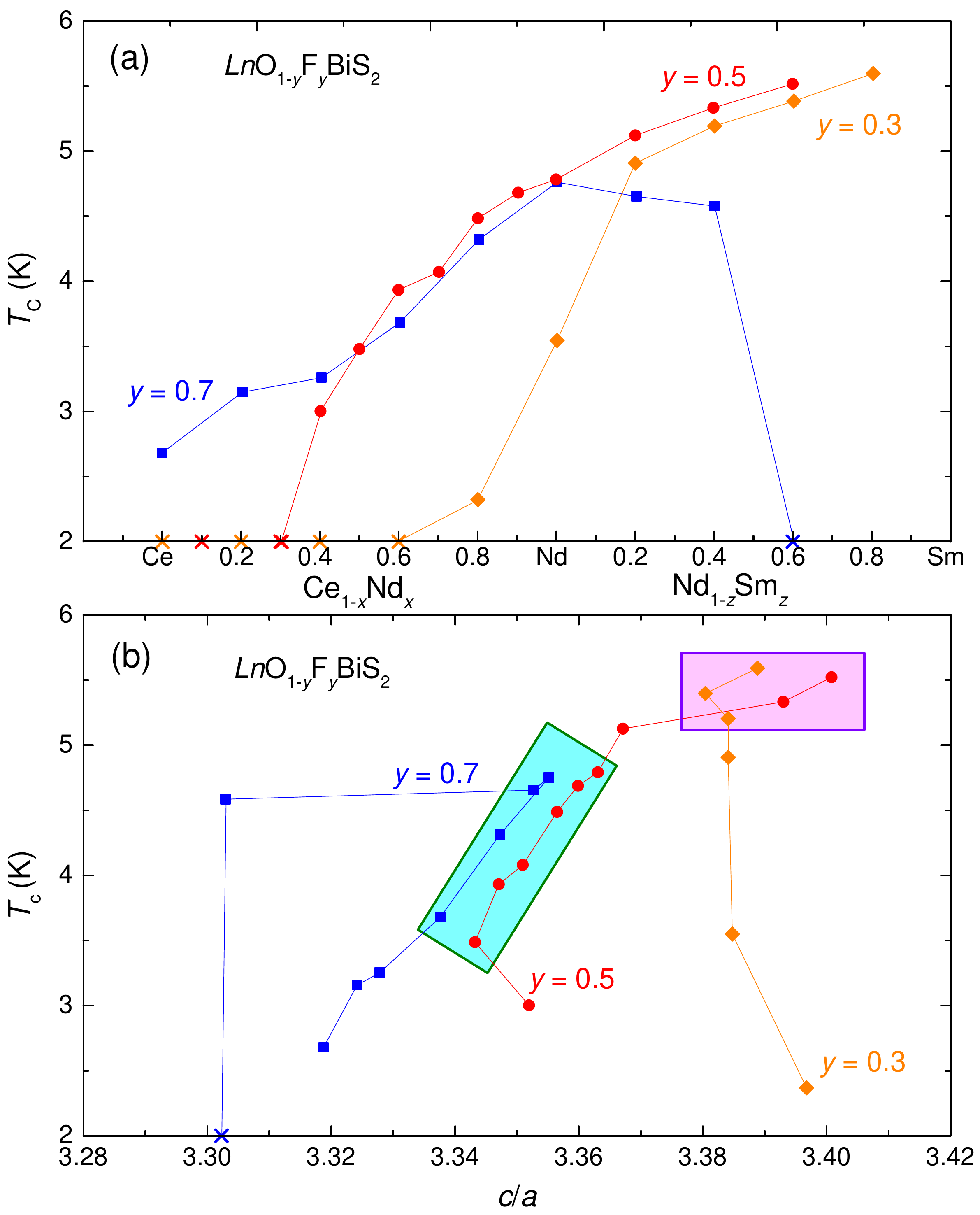}
  \caption{(Color online) (Color online) Superconducting critical temperature $T_c$ of \textit{Ln}O$_{1-y}$F$_{y}$BiS$_{2}$ ($y$ = 0.7, 0.5, and 0.3) as a function of \textit{Ln} concentration. The cross symbols indicate that a superconducting transition is not observed above 2 K. Reprinted with permission from Ref. \cite{kajitani2015chemical} J. Kajitani, T. Hiroi, A. Omachi, O. Miura, Y. Mizuguchi, J. Phys. Soc. Jpn. 84 (2015) 044712. Copyrighted by the Physical Society of Japan.}
\label{FIG.14.}
\end{figure}

The role of the charge carrier density in the enhancement of superconductivity is emphasized in the study of the La$_{1-x}$Mg$_{x}$O$_{1-2x}$F$_{2x}$BiS$_{2}$ system in which  Mg$^{2+}$ and F$^{-}$ are substituted for La$^{3+}$ and O$^{2-}$, respectively \cite{chen2013effect}. Superconductivity is observed for nominal $x$ $\geqslant$ 0.2 where $T_c$ continues to increase up to $\sim$3 K at $x$ = 0.3 which is slightly beyond the solubility limit of $x$ = 0.25. For a fixed F concentration at 2$x$ = 0.4, the introduction of Mg$^{2+}$ ions into the LaO$_{0.6}$F$_{0.4}$BiS$_{2}$ compound generates holes in the BiS$_2$ superconducting layers which have the effect of suppressing superconductivity and thereby reducing $T_c$ despite the reduction in lattice parameters. The research also indicates that a reduction in lattice parameters leads to an enhancement of $T_c$. For instance, a reduction in the values of the lattice parameters appears to enhance superconductivity when comparing the La$_{0.8}$Ca$_{0.2}$O$_{0.6}$F$_{0.4}$BiS$_{2}$ and La$_{0.8}$Mg$_{0.2}$O$_{0.6}$F$_{0.4}$BiS$_{2}$ compounds which have $T_c$ values of 2.07 K and 2.54 K, respectively.

\section{Superconductivity of \textit{Ln}O$_{0.5}$F$_{0.5}$BiS$_{2}$ under applied pressure}

In addition to chemical substitution, the application of external pressure is another means to induce fundamental change to both the crystalline and electronic structure which can have a significant impact on the superconducting properties of the layered BiS$_2$-based compounds. The first report of a dramatic and abrupt increase in $T_c$ from $\sim$3 to 10.7 K for the LaO$_{0.5}$F$_{0.5}$BiS$_2$ compound under pressure is shown in the upper panel of Fig. 15(a). In the $T_c$($P$) phase diagram, there is an obvious transition from a low-$T_c$ superconducting phase to a high-$T_c$ superconducting phase \cite{wolowiec2013pressure}. The phenomenon was soon reported to be associated with a pressure-induced first order phase transition from a tetragonal to a monoclinic crystal structure in which there is a relative shift or sliding of the two neighboring BiS$_2$ layers along the $a$ axis \cite{tomita2014pressure}. Similar  behavior in the evolution of $T_c$ with pressure has also been observed in other \textit{Ln}O$_{0.5}$F$_{0.5}$BiS$_{2}$ (\textit{Ln} = Ce, Pr, Nd) compounds as shown in the $T_c$($P$) phase diagrams of Fig. 16 \cite{wolowiec2013}.

\begin{figure}[h]
\centering
\includegraphics[width=7.5cm]{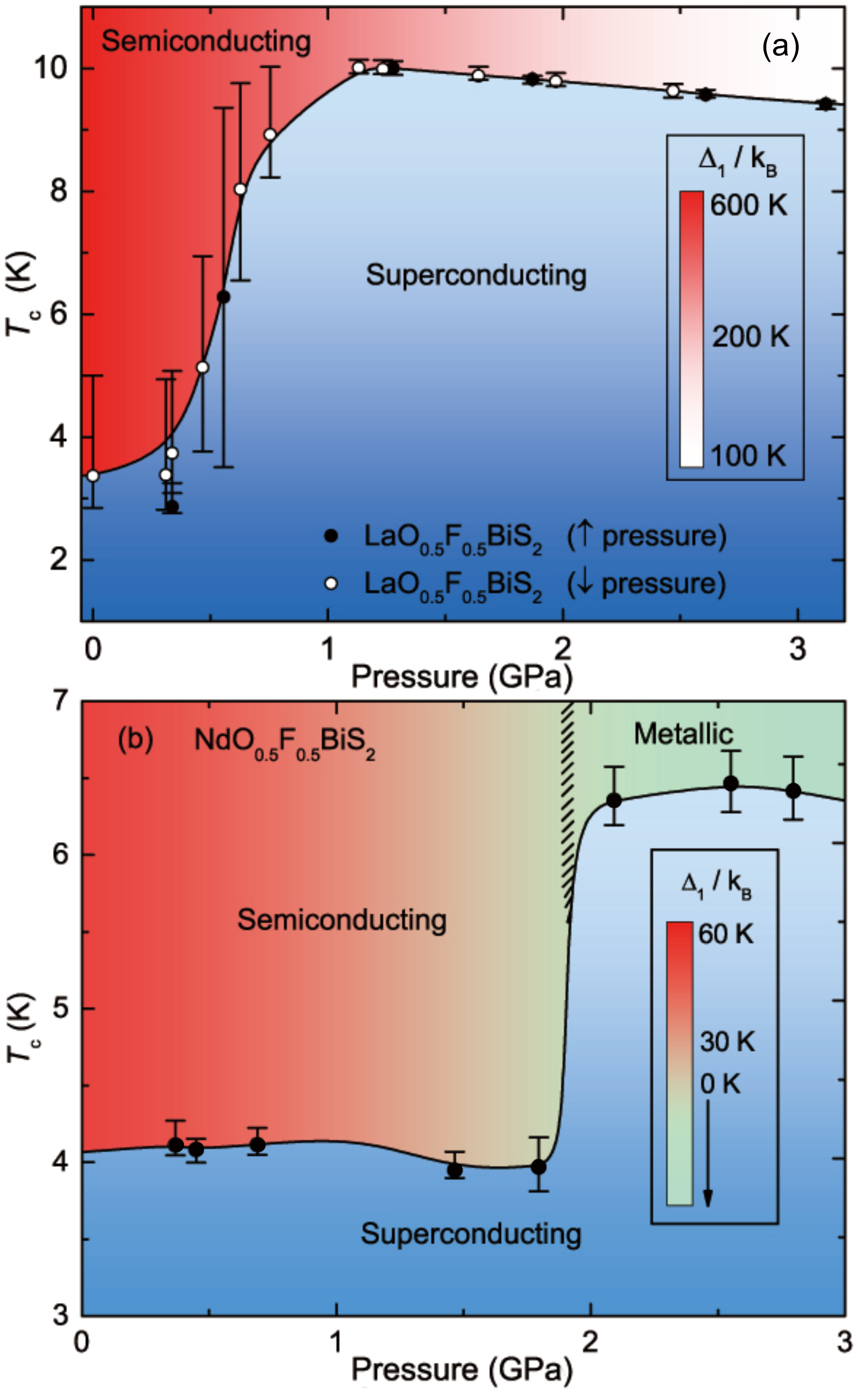}
  \caption{(Color online) $T_c$($P$) phase diagrams for (a) LaO$_{0.5}$F$_{0.5}$BiS$_{2}$ and (b) NdO$_{0.5}$F$_{0.5}$BiS$_2$. The red color gradient represents the suppression of the semiconducting behavior with pressure as manifested in the decrease of the semiconducting energy gap $\Delta$$_1$/k$_{\rm B}$ whose values are indicated in the legend. The green region to the right of the crosshatching in (b) corresponds to the metallization of NdO$_{0.5}$F$_{0.5}$BiS$_2$. (a), (b) Reprinted with permission from Ref. \cite{wolowiec2013pressure} (copyright (2013) American Physical Society) and Ref. \cite{wolowiec2013} (Enhancement of superconductivity near the pressure-induced semiconductor-metal transition in the BiS$_2$-based superconductors \textit{Ln}O$_{0.5}$F$_{0.5}$BiS$_2$ (\textit{Ln} = La, Ce, Pr, Nd), J. Phys.: Condens. Matter 25 (2013) 422201. Copyright (2013) by the IOP Publishing.), respectively. For interpretation of the definition of $T_c$, vertical bars, and $\Delta$$_1$/k$_{\rm B}$, the reader is referred to the web versions of the corresponding articles.}
\label{FIG.15.}
\end{figure}

Interestingly, the critical transition pressure, $P_{\rm t}$, that marks the low-$T_c$ to high-$T_c$ phase transition in the superconducting state is coincident with changes in the normal state electrical transport properties in the \textit{Ln}O$_{0.5}$F$_{0.5}$BiS$_{2}$ system; the pressure-induced phase transition corresponds with a saturation in the rate of suppression of semiconducting-like behavior in the normal state \cite{wolowiec2013pressure}. In the case of the NdO$_{0.5}$F$_{0.5}$BiS$_2$ compound, a semiconductor-metal transition occurs at $P_{\rm t}$ as shown in Fig. 15(b) \cite{wolowiec2013}. Such changes in the behavior of the normal state $\rho$($T$) indicate that there may be increases in the density of charge carriers that correspond to the rapid increase in $T_c$.


AG samples from the series of \textit{Ln}O$_{0.5}$F$_{0.5}$BiS$_{2}$ compounds with \textit{Ln} elements of smaller atomic radius (i.e., larger atomic number) exhibit higher $T_c$ values when compared with compounds containing \textit{Ln} elements of larger atomic radius suggesting that $T_c$ correlates with the lattice parameters. Given the correlation between $T_c$ and the lattice parameters, one might expect that the pressure-induced phase transitions observed for the various \textit{Ln}O$_{0.5}$F$_{0.5}$BiS$_{2}$ compounds would occur at a characteristic and consistent value of the lattice parameter $a$ across the \textit{Ln}O$_{0.5}$F$_{0.5}$BiS$_{2}$ series. Furthermore, considering the results from substitution studies discussed earlier in this article, one would expect a smaller applied pressure to induce the phase transition for those \textit{Ln}O$_{0.5}$F$_{0.5}$BiS$_{2}$ compounds with \textit{Ln} elements exhibiting smaller values of the lattice parameter $a$. To the contrary, the $a$-axis lattice parameters at $P_{\rm t}$ are not constant across the series \textit{Ln}O$_{0.5}$F$_{0.5}$BiS$_{2}$. The lattice parameter $a$ for the LaO$_{0.5}$F$_{0.5}$BiS$_{2}$ compound just before the transition pressure, \textit{P$_{\rm t}$} $\sim$ 0.7 GPa, is estimated to be 4.08 $\AA$ which is larger than the values of $a$ for other members of the \textit{Ln}O$_{0.5}$F$_{0.5}$BiS$_{2}$ series at ambient pressure \cite{tomita2014pressure,yazici2013superconductivity}. The $T_c$($P$) phase diagrams depicted in Fig. 16 indicate that the critical transition pressure $P_{\rm t}$ increases as the atomic size of the \textit{Ln} ions decreases and that the jump in $T_c$ during the phase transition ($\Updelta$$T_c$) is smaller for smaller \textit{Ln} ions in \textit{Ln}O$_{0.5}$F$_{0.5}$BiS$_{2}$ \cite{wolowiec2013pressure,wolowiec2013}. Heavier \textit{Ln} ions seem to help stabilize the structure of the \textit{Ln}O$_{0.5}$F$_{0.5}$BiS$_{2}$ against external pressure.

\begin{figure}[t]
\centering
\includegraphics[width=8cm]{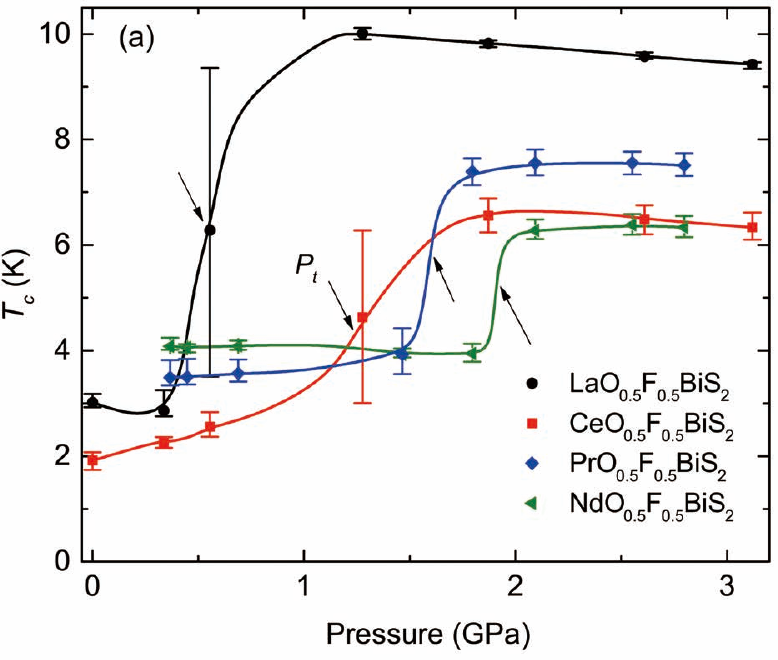}
  \caption{(Color online) Pressure dependence of $T_c$ for \textit{Ln}O$_{0.5}$F$_{0.5}$BiS$_{2}$ (\textit{Ln} = La, Ce, Pr, Nd). The black arrows indicate the critical transition pressures from the low-$T_c$ to high-$T_c$ superconducting phases. Reprinted with permission from Ref. \cite{wolowiec2013}: Enhancement of superconductivity near the pressure-induced semiconductor-metal transition in the BiS$_2$-based superconductors \textit{Ln}O$_{0.5}$F$_{0.5}$BiS$_2$ (\textit{Ln} = La, Ce, Pr, Nd), J. Phys.: Condens. Matter 25 (2013) 422201. Copyright (2013) by the IOP Publishing.}
\label{FIG.16.}
\end{figure}

Figure 17 shows the evolution of $T_c$ for the La$_{1-x}$Sm$_{x}$O$_{0.5}$F$_{0.5}$BiS$_{2}$ compound as a function of both pressure and  Sm concentration $x$ \cite{fang2015unpublished}. Similar to the \textit{Ln}O$_{0.5}$F$_{0.5}$BiS$_{2}$ compounds, the La$_{1-x}$Sm$_{x}$O$_{0.5}$F$_{0.5}$BiS$_{2}$ system exhibits a reversible pressure-induced transition from a low-$T_c$ (SC1) to a high-$T_c$ (SC2) superconducting phase. It has been shown that the lattice parameter $a$ in the  La$_{1-x}$Sm$_{x}$O$_{0.5}$F$_{0.5}$BiS$_{2}$ compound is significantly reduced by increasing the Sm concentration $x$ \cite{fang2015enhancement}. The results for the SC1 to SC2 phase transition in the La$_{1-x}$Sm$_{x}$O$_{0.5}$F$_{0.5}$BiS$_{2}$ system indicate that optimization of $T_c$ could be achieved in La$_{1-x}$Sm$_{x}$O$_{0.5}$F$_{0.5}$BiS$_{2}$ by decreasing the $a$ lattice parameter (high Sm concentration) in the SC1 phase at ambient pressure or by increasing $a$ (low Sm concentration) in the SC2 phase under pressure.


\begin{figure}[t]
\centering
\includegraphics[width=8cm]{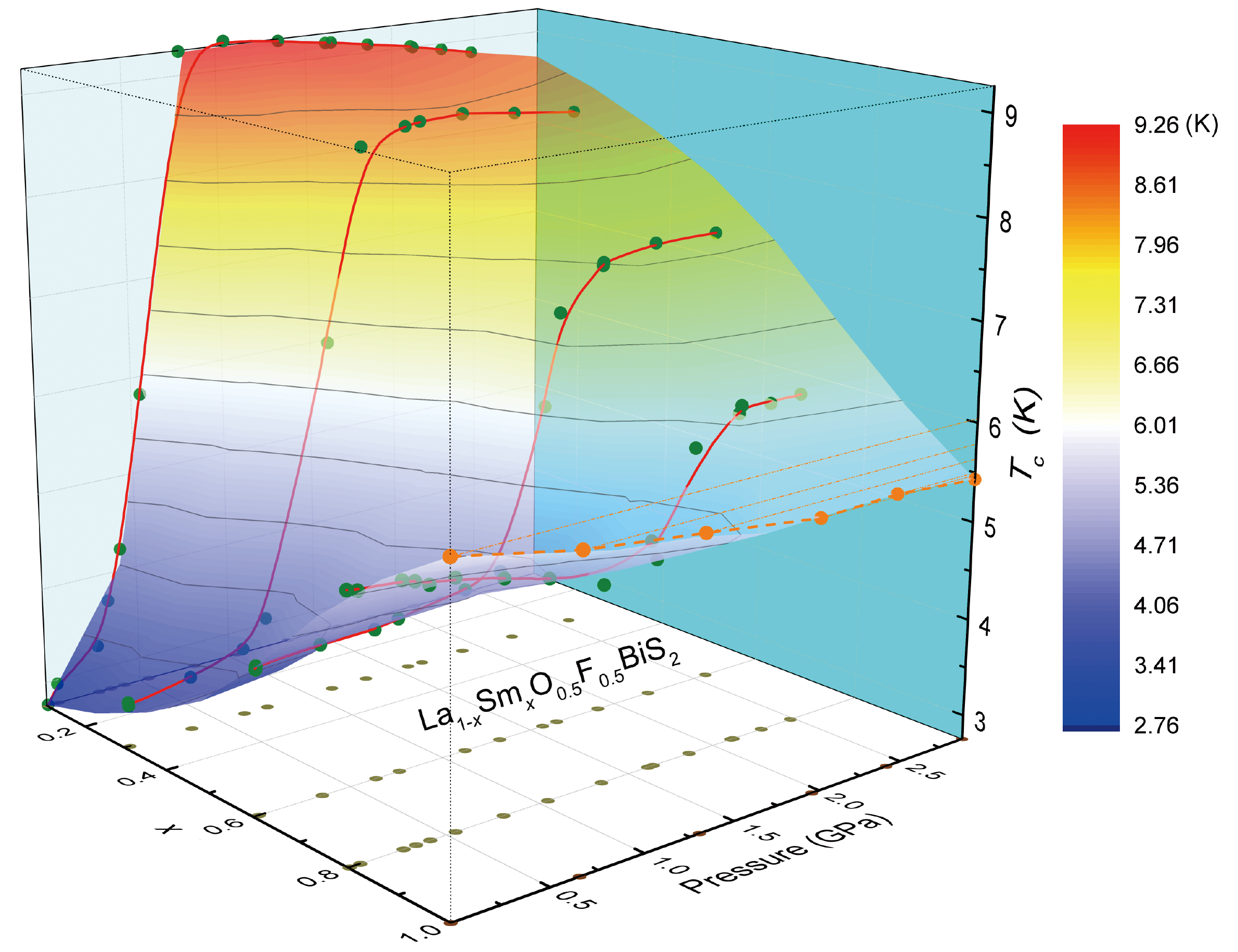}
  \caption{(Color online) Superconducting critical temperature $T_c$ of La$_{1-x}$Sm$_{x}$O$_{0.5}$F$_{0.5}$BiS$_{2}$ ($x$ = 0.1-0.8) plotted as a function of pressure and nominal Sm concentration $x$. Filled circles in the $x$-$y$ plane are projections of $T_c$. The data are taken from Ref. \cite{fang2015unpublished}. (For interpretation of the estimated $T_c$ values of SmO$_{0.5}$F$_{0.5}$BiS$_2$, the reader is referred to Ref. \cite{fang2015unpublished}.) }
\label{FIG.17.}
\end{figure}

Although the chemical composition of the blocking layers in the \textit{Ln}O$_{0.5}$F$_{0.5}$BiS$_{2}$ and La$_{1-x}$Sm$_{x}$O$_{0.5}$F$_{0.5}$BiS$_{2}$ compounds are different, their evolution of $T_c$ with pressure is similar. As displayed in Fig. 18(a), $P_{\rm t}$ increases with decreasing atomic size of the \textit{Ln} ion for \textit{Ln}O$_{0.5}$F$_{0.5}$BiS$_{2}$ (\textit{Ln} = La, Ce, Pr, Nd) compounds in the same way that $P_{\rm t}$ increases with increasing Sm concentration for La$_{1-x}$Sm$_{x}$O$_{0.5}$F$_{0.5}$BiS$_{2}$ under pressure \cite{wolowiec2013,fang2015unpublished}. If we plot $\Updelta$$T_c$ for \textit{Ln}O$_{0.5}$F$_{0.5}$BiS$_{2}$ and La$_{1-x}$Sm$_{x}$O$_{0.5}$F$_{0.5}$BiS$_{2}$ as a function of lattice parameter $a$, they follow the same linear trend, despite the differences in the chemical composition of their blocking layer.  This suggests the importance of the in-plane separation between ions in the superconducting BiS$_2$ layers and its effect on superconductivity in the BiS$_2$-based compounds under pressure. However, there is no clear relationship observed between the $c$ lattice parameter and superconductivity in these BiS$_2$-based compounds under pressure.




\begin{figure}[t]
\centering
\includegraphics[width=8cm]{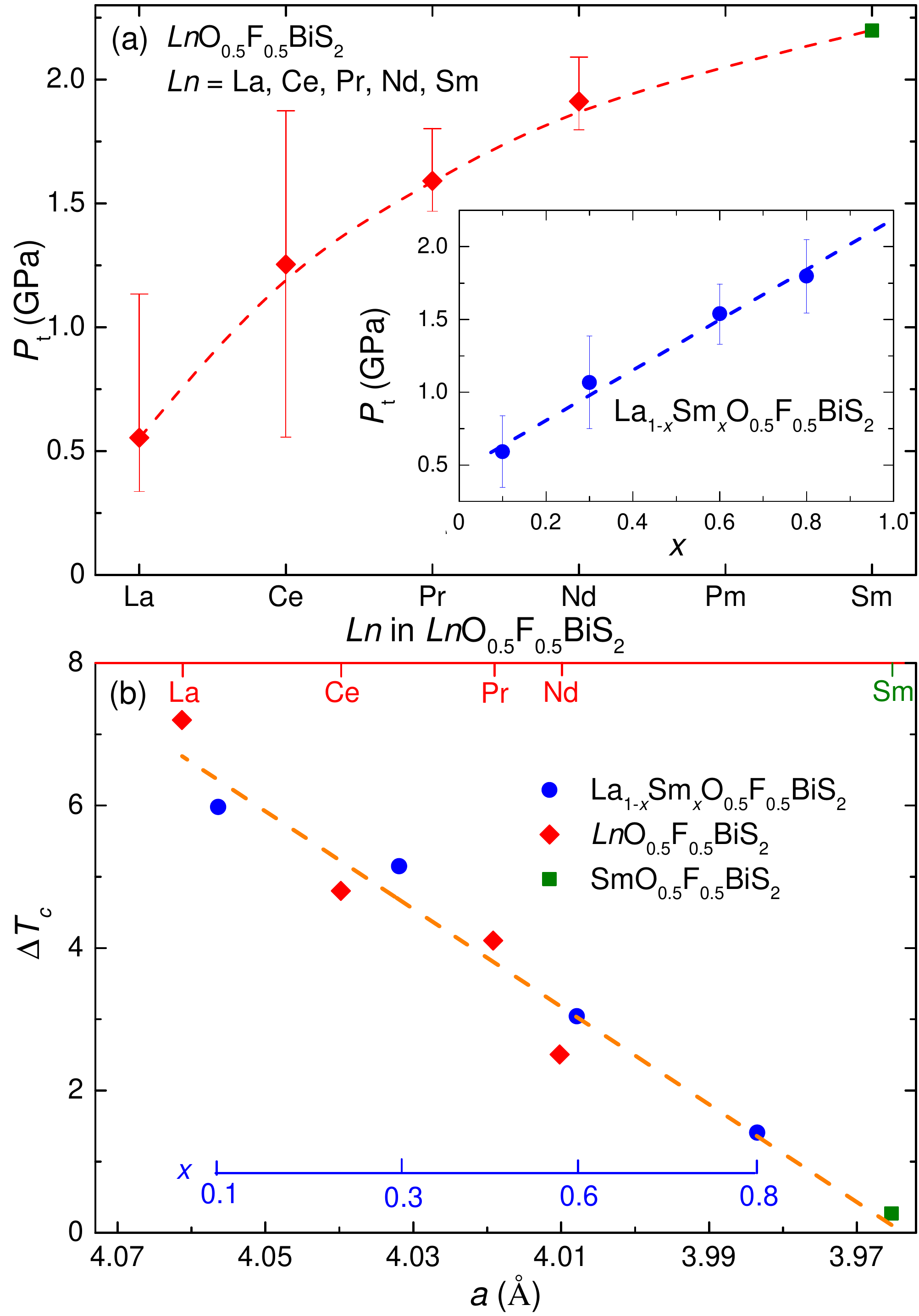}
  \caption{(Color online) (a) Evolution of $P_{\rm t}$ as a function of \textit{Ln} in \textit{Ln}O$_{0.5}$F$_{0.5}$BiS$_2$. The inset in panel (a) shows Sm concentration dependence of $P_{\rm t}$ for La$_{1-x}$Sm$_{x}$O$_{0.5}$F$_{0.5}$BiS$_{2}$.  (b) $\Delta$$T_c$ vs lattice parameter $a$ of La$_{1-x}$Sm$_{x}$O$_{0.5}$F$_{0.5}$BiS$_{2}$ and \textit{Ln}O$_{0.5}$F$_{0.5}$BiS$_{2}$. Dashed lines are guides to the eye. $\Delta$$T_c$ and $P_{\rm t}$ values of \textit{Ln}O$_{0.5}$F$_{0.5}$BiS$_2$ and La$_{1-x}$Sm$_{x}$O$_{0.5}$F$_{0.5}$BiS$_2$ are taken from Refs. \cite{wolowiec2013} and \cite{fang2015unpublished}, respectively. The values of the lattice parameter $a$ for \textit{Ln}O$_{0.5}$F$_{0.5}$BiS$_2$ and La$_{1-x}$Sm$_{x}$O$_{0.5}$F$_{0.5}$BiS$_2$ are obtained from Refs. \cite{yazici2015superconductivity} and \cite{fang2015enhancement}, respectively. (For interpretation of the estimated data of SmO$_{0.5}$F$_{0.5}$BiS$_2$, the reader is referred to Refs. \cite{fang2015enhancement,fang2015unpublished}.)}
\label{FIG.18.}
\end{figure}

Partial substitution of \textit{Ln} ions with smaller ions typically results in a reduction of the $a$ lattice parameter and along with (but not always) a slight increase in the $c$ lattice parameter; however, it has been reported that both the $a$ and $c$ lattice parameters can be simultaneously suppressed with the application of hydrostatic pressure \cite{tomita2014pressure}. This has implications for the behavior of superconductivity under external pressure as being fundamentally different when compared to the behavior of superconductivity in response to chemical substitution in the BiS$_2$-based compounds. As an example, it was reported that a structural phase transition from the tetragonal phase to the monoclinic phase occurs at $\sim$0.7 GPa in LaO$_{0.5}$F$_{0.5}$BiS$_{2}$ under pressure, whereas it is known that all BiS$_2$-based superconductors at ambient pressure exhibit the tetragonal phase, including those compounds subject to heavy chemical substitution. As another example, the $T_c$ of AG samples of LaO$_{0.5}$F$_{0.5}$BiS$_{2}$ increases with a reduction in the lattice parameter $a$ through a partial or full replacement of La with heavier (smaller) \textit{Ln}, whereas further reduction of the lattice parameter $a$ by increasing external hydrostatic pressure up to 1.5 GPa results in a gradual decrease in the $T_c$ of La$_{0.2}$Sm$_{0.8}$O$_{0.5}$F$_{0.5}$BiS$_{2}$ \cite{fang2015unpublished}. Rather than adopting a single lattice parameter determination in explaining the effects of applied pressure on superconductivity in these BiS$_2$-based compounds, these observations suggest that there is a complex interplay of various parameters and their effects on superconductivity under pressure. In addition to the importance of the size of the lattice constant $a$ in affecting superconductivity, reduction of the lattice constant $c$ also appears to affect the superconducting state of BiS$_2$-based compounds. By applying uniaxial pressure along the $c$-axis in single crystalline PrO$_{0.5}$F$_{0.5}$BiS$_2$, a low-$T_c$ to high-$T_c$ phase transition is also observed at 0.7 GPa, which is significantly lower than the $P_{\rm t}$ of PrO$_{0.5}$F$_{0.5}$BiS$_2$ under hydrostatic pressure \cite{fujioka2014high}. Until now, although a large number of compounds in the BiS$_{2}$ family have been investigated in terms of their transport behavior under pressure, reports on the evolution of crystal structure with pressure are very limited. To clarify correlations between the crystal structure of BiS$_{2}$-based compounds and their superconductivity under pressure, further experiments involving the use of in-situ x-ray diffraction/absorption measurements under pressure need to be performed on different BiS$_2$-based samples in both polycrystal and single crystal form.



\section{Se-substituted \textit{Ln}O$_{0.5}$F$_{0.5}$BiS$_{2}$}

Most of the research regarding chemical substitution and superconductivity in the \textit{Ln}(O,F)BiS$_2$ compounds has been devoted to studying the effects of elemental substitution in the blocking \textit{Ln}O layers. There has been much less effort and only a few reports regarding the effects on superconductivity that result from modification of the superconducting BiS$_2$ layers. Substitution of Sb for Bi appears to be an ineffective means for enhancing superconductivity. While LaO$_{0.5}$F$_{0.5}$BiS$_2$ has a $T_c$ $\sim$ 3 K, LaO$_{0.5}$F$_{0.5}$SbS$_2$ does not exhibit superconductivity above 1.7 K \cite{krzton2014superconductivity}. Furthermore, it appears that a small amount of Sb substitution (10$\%$) in the NdO$_{0.5}$F$_{0.5}$BiS$_2$ compound strongly suppresses superconductivity \cite{hiroi2015element}. Alternatively, it is possible to completely replace S with either Se or Te in LaO$_{0.5}$F$_{0.5}$BiS$_2$; however, only the LaO$_{0.5}$F$_{0.5}$BiSe$_2$ compound was found to exhibit superconductivity with a $T_c$ $\sim$ 2.6 K \cite{krzton2014superconductivity}. The Se$^{2-}$ and S$^{2-}$ ions are chemically similar; however, the Se$^{2-}$ ionic radius of 1.98 $\AA$ is larger than the S$^{2-}$ ionic radius of 1.84 $\AA$ (assuming a coordination number of 6) \cite{mizuguchi2015in}. This distinction may prove useful in studies on the effects of Se substitution in the \textit{Ln}O$_{0.5}$F$_{0.5}$BiS$_2$ system from which new insights and perspectives might be gained regarding the relationship between crystal structure and superconductivity in the \textit{Ln}(O,F)BiS$_2$ compounds.

\begin{figure}[t]
\centering
\includegraphics[width=8cm]{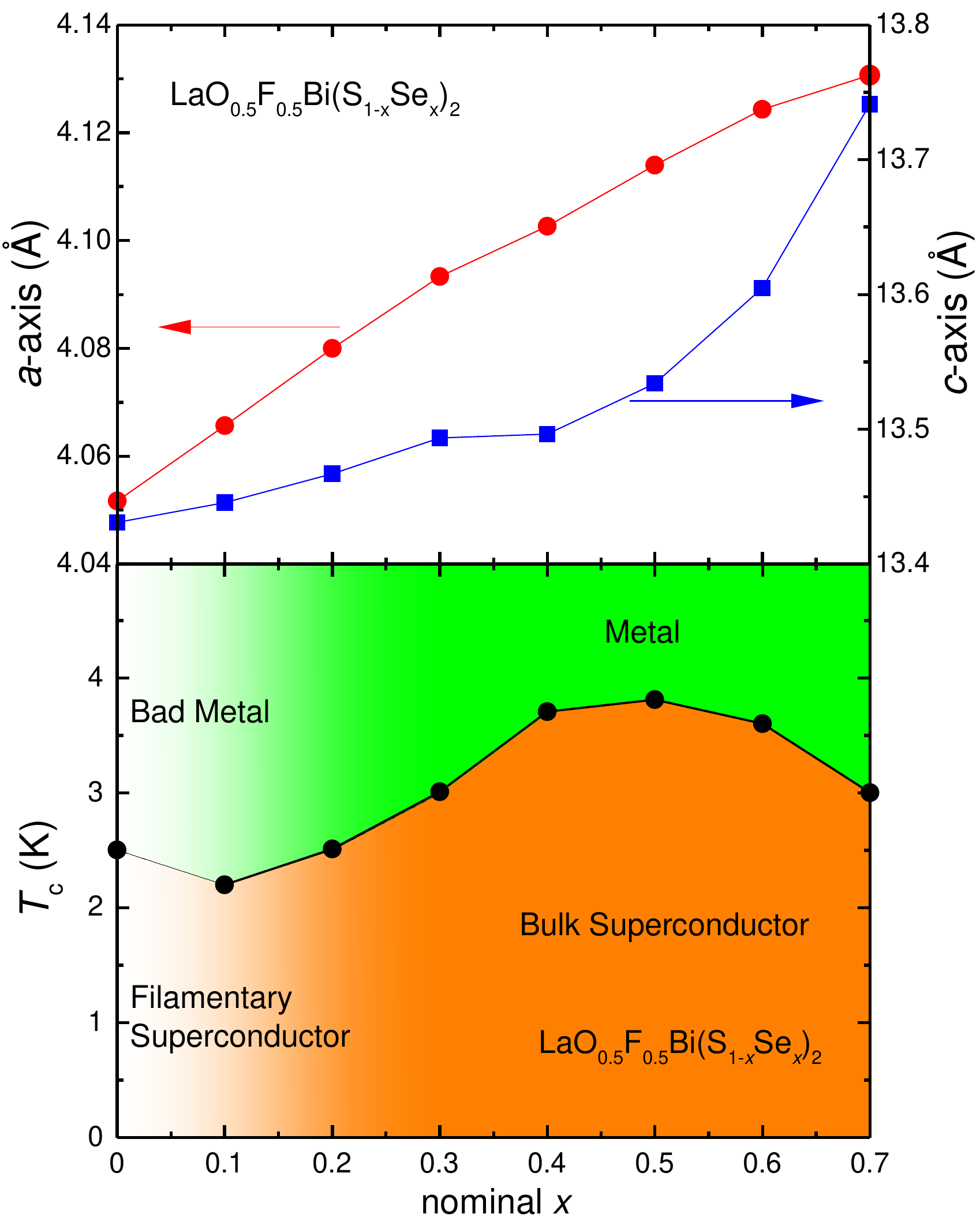}
 \caption{(Color online) Nominal Se concentration dependence of lattice constants $a$ and $c$ (upper) and $T_c$ (lower). Reprinted with permission from Ref. \cite{hiroi2015evolution} T. Hiroi, J. Kajitani, A. Omachi, O. Miura, Y. Mizuguchi, J. Phys. Soc. Jpn. 84 (2015) 024723. Copyrighted by the Physical Society of Japan.}
\label{FIG.19.}
\end{figure}

The dependence of $T_c$ on the concentration of Se, substituted for S in \textit{Ln}O$_{0.5}$F$_{0.5}$BiS$_{2}$, exhibits complicated behavior when the \textit{Ln} element changes. For polycrystalline  NdO$_{0.5}$F$_{0.5}$Bi(S$_{1-x}$Se$_{x}$)$_{2}$ ($x$ $\leqslant$ 0.2), it has been reported that, as the Se concentration is increased, there is a noticeable change in the lattice constant $a$, but no significant change in the lattice constant $c$; in addition, both $T_c$ and the shielding volume fraction of NdO$_{0.5}$F$_{0.5}$Bi(S$_{1-x}$Se$_{x}$)$_{2}$ decrease with increasing Se concentration \cite{hiroi2015element}. These results are consistent with the previous discussion regarding modification of the \textit{Ln}O blocking layers in which superconductivity is enhanced as the lattice parameters are reduced. In the case of LaO$_{0.5}$F$_{0.5}$Bi(S$_{1-x}$Se$_x$)$_2$, $T_c$ increases from 2.4 K at $x$ = 0.2 to 3.8 K at $x$ = 0.5 with a large shielding volume fraction followed by a slight decrease in $T_c$ for $x$ $\textgreater$ 0.5 \cite{mizuguchi2015in,hiroi2015evolution}. Synchrotron x-ray diffraction experiments performed on the LaO$_{0.5}$F$_{0.5}$Bi(S$_{1-x}$Se$_{x}$)$_{2}$ compound show that Se atoms are more likely to occupy in-plane S1 sites rather than the out-of-plane S2 sites, resulting in an expansion in both the $a$ and $c$ axes with increasing Se concentration \cite{mizuguchi2015in}. This behavior complicates the picture we have regarding the correlation between the crystal structure and superconductivity in the BiS$_2$-based compounds.

\section{Concluding remarks}

There is already a significant amount of interesting behavior reported on the effects of chemical substitution in the recently discovered family of BiS$_2$-based compounds; and many of these interesting properties are often not exhibited in the parent form of these compounds. The \textit{Ln}BiOS$_2$ compounds were synthesized and reported as early as 1995 \cite{tanryverdiev1995synthesis}; however, no superconductivity was found in the \textit{Ln}BiOS$_{2}$ system until electron doping studies were performed in 2012 \cite{demura2013new,mizuguchi2012superconductivity,xing2012superconductivity,yazici2013superconductivity}. Hence, the introduction of charge carriers (electrons) into the parent compound is regarded to be essential for the emergence of superconductivity in BiS$_2$-based superconductors \cite{yazici2015superconductivity,mizuguchi2012superconductivity}. Measurements of $\rho$($T$) under pressure provide supporting evidence that the dramatic enhancement  of superconductivity in various \textit{Ln}O$_{0.5}$F$_{0.5}$BiS$_2$ compounds is coincident with an increase in the charge carrier density \cite{wolowiec2013,wolowiec2013pressure}.

Substitution for ions in either the BiS$_2$ superconducting layers or the blocking \textit{Ln}O layers with ions of different atomic size allows one to tune the lattice parameters as well as the local lattice structure of the BiS$_2$-based compounds and thereby generate chemical pressure. An increase in chemical pressure in the superconducting BiS$_2$ layers would result in a shorter Bi-S1 bond distance thereby enhancing the overlap of the Bi-6p and S-p orbitals; this would allow superconductivity to be tuned without a dramatic change of charge carrier density. Experimental evidence suggests that both the contraction along the $a$ axis and the $c$/$a$ ratio play an essential role in the emergence of superconductivity in some \textit{Ln}Bi(O,F)S$_2$ compounds \cite{fang2015enhancement,kajitani2014chemical,kajitani2015chemical,yazici2013superconductivity,chen2013effect}. However, the clear correlation between lattice contraction and superconductivity does not apply to all BiS$_2$-based compounds and fails to hold when samples are subjected to hydrostatic pressure. At present, finding a universal relationship between the crystal structure and superconductivity in the BiS$_{2}$-based compounds is still one of the main challenges in this research area. Recently, Mizuguchi \textit{et.al.} defined a value of in-plane chemical pressure as (R$_{\rm Bi}$ + R$_{\rm S1}$)/Bi-S1(in-plane), where R$_{Bi}$ is the estimated ionic radius of Bi$^{2.5+}$, R$_{S1}$ is the estimated average ionic radius of the chalcogen at the S1 site, and the Bi-S1 bond distance is obtained from the Rietveld refinement of experimental XRD data \cite{mizuguchi2015in}. A clear correlation between $T_c$ and the in-plane chemical pressure is observed in the \textit{Ln}O$_{0.5}$F$_{0.5}$BiS$_{2}$ (\textit{Ln} = Ce$_{1-x}$Nd$_{x}$, Nd$_{1-x}$Sm$_{x}$) and LaO$_{0.5}$F$_{0.5}$Bi(S$_{1-x}$Se$_{x}$)$_{2}$ systems; however, such a relationship needs to be further confirmed with studies of the effects of in-plane chemical pressure on other BiS$_2$-based systems. Additional experimental and theoretical effects are also needed to elucidate the relationship between in-plane chemical pressure and the resultant enhancement of $T_c$ \cite{mizuguchi2015in}.

At present, the highest reported $T_c$ among the BiS$_2$-based compounds is $\sim$10.7 K for LaO$_{0.5}$F$_{0.5}$BiS$_2$ under pressure \cite{wolowiec2013pressure,tomita2014pressure}; however, the BSCCO cuprate superconductors, in which Bi ions are located in the blocking layers of the structures, can have $T_c$ values higher than 100 K \cite{maeda1988new,subramanian1988new,tallon1988high}. The Ba$_{1-x}$K$_{x}$BiO$_3$ family, another system containing Bi ions, was reported to have $T_c$ values as high as $\sim$30 K and the BiO$_2$ planes are regarded to be essential for the superconductivity in the compounds \cite{cava1988superconductivity,mattheiss1988electronic}. The speed at which new members of BiS$_2$-based superconductors are being synthesized and studied holds promise for the discovery of new superconducting compounds with values of $T_c$ significantly higher than 11 K. Nevertheless, some challenging but essential problems regarding the intrinsic behavior of the BiS$_{2}$-based superconductors remain to be solved. As the debate over the nature of the pairing mechanism for superconductivity in the BiS$_{2}$-based superconductors continues, it is difficult to develop an effective strategy for enhancing $T_c$.  The intrinsic transport and thermal properties, which may shed light on the nature of the pairing mechanism, need to be established through further studies on high quality single-crystalline samples. Theoretical calculations, which investigate electronic structure and related properties of BiS$_2$-based superconductors, should be more focused on the actual chemical compositions of the compounds. Another interesting but unclarified issue is the remarkable difference in $T_c$ between the AG  and HP annealed samples of \textit{Ln}(O,F)BiS$_{2}$ (\textit{Ln} = La, Ce, Pr). The similarity in both chemical composition and crystal structure of the HP annealed samples and their AG counterparts suggests that the local environment of the crystals may play an essential role in the superconductivity of these BiS$_{2}$-based compounds. It is hopeful that $T_c$ values for the family of BiS$_2$-based superconductors can be further enhanced if the factors responsible for an increase in $T_c$ such as those affected by HP annealing and the application of external pressure are clarified. Amidst the abundance of physical phenomena thus far observed, at this early stage of discovery and research there remain unanswered questions and opportunities to develop a fundamental understanding of superconductivity in the BiS$_2$-based compounds.


\section{Acknowledgements}

Preparation of this review was supported by the U.S. Department of Energy, Office of Basic Energy Sciences, Division of Materials Sciences and Engineering under Award Grant No. DE-FG02-04-ER46105, the National Nuclear Security Administration under the Stewardship Science Academic Alliance program through the U.S. Department of Energy Grant No. DE-NA0001841, and the National Science Foundation under Grant No. DMR-1206553. Helpful discussions with B. D. White are gratefully acknowledged.


\clearpage

\end{document}